\documentclass[12pt]{article}
\usepackage{graphicx}
\usepackage{epsfig}
\usepackage{amssymb}
\usepackage{natbib}

\usepackage{epsfig}
\usepackage{graphics}
\usepackage{longtable}
\usepackage{graphicx}

\usepackage{float}
\usepackage{txfonts}
\usepackage{pslatex}

\begin{document}

\begin{center}
\vbox to 1 truecm {}

{\large \bf First performance studies of a prototype for the CASTOR }\par 
\vskip 3 truemm
{\large \bf forward calorimeter at the CMS experiment} \\
\vskip 1. truecm

{\bf X.~Aslanoglou$^1$, A.~Cyz$^2$, N.~Davis$^3$, D.~d'Enterria$^4$, E.~Gladysz-Dziadus$^2$, 
C.~Kalfas$^5$, Y.~Musienko$^6$, A.~Kuznetsov$^6$, A.~D.~Panagiotou$^3$}\\
\vskip 8 truemm

{\it $^1$ University of Ioannina, PO Box 1186, 45110 Ionnanina, Hellas}
\vskip 3 truemm
{\it $^2$ Institute of Nuclear Physics, Radzikowskiego 152, 31342 Krakow, Poland}
\vskip 3 truemm
{\it $^3$ University of Athens, Phys. Dept. 15701 Athens, Hellas}
\vskip 3 truemm
{\it $^4$ CERN, PH Dept., 1211 Geneva 23, Switzerland}
\vskip 3 truemm
{\it $^5$ NRC ``Demokritos'' INP, PO Box 60228, 15310 Ionnanina, Hellas}
\vskip 3 truemm
{\it $^6$ Northeastern University, Dept. of Physics, Boston, MA 02215, USA}
\vskip 3 truemm
\end{center}

\begin{abstract}
We present results on the performance of the first prototype of the 
CASTOR quartz-tungsten sampling calorimeter, to be installed in the very forward region 
of the CMS experiment at the LHC. This study includes {\sc geant} Monte Carlo simulations 
of the \v{C}erenkov light transmission efficiency of different types of air-core 
light guides, as well as analysis of the calorimeter linearity and resolution 
as a function of energy and impact-point, obtained with 20-200 GeV electron beams 
from CERN/SPS tests in 2003. Several configurations of the calorimeter have been 
tested and compared, including different combinations of (i) structures for the 
active material of the calorimeter (quartz plates and fibres), (ii) various light-guide 
reflecting materials (glass and foil reflectors) and (iii) photodetector devices 
(photomultipliers and avalanche photodiodes).
\end{abstract}



\section{Introduction}

The CASTOR (Centauro And Strange Object Research) detector is a quartz-tungsten 
sampling calorimeter that has been proposed to study the very forward rapidity 
(baryon-rich) region in heavy ion collisions in the multi-TeV range at the LHC~\cite{castor1} 
and thus to complement the heavy ion physics programme, focused mainly in the baryon-free 
midrapidity region~\cite{ptdr}. CASTOR will be installed in the CMS experiment at 14.38 m from the 
interaction point, covering the pseudorapidity range 5.2 $< \eta <$ 6.6 and will, thus, 
contribute not only to the heavy ion program, but also to diffractive and low-$x$ physics 
in pp collisions~\cite{castor_forw}. The CMS and TOTEM experiments supplemented by the 
CASTOR detector will constitute the largest acceptance system ever built at a hadron 
collider, having the possibility to measure the forward energy and particle flow up to 
$\eta$ = 6.6. With the design specifications for CASTOR,
the total and the electromagnetic energies in its acceptance range  ($E_{tot}\sim$180 TeV 
and $E_{em}\sim$50 TeV respectively according to {\sc hijing}~\cite{hijing} PbPb simulations 
at 5.5 TeV) can be measured with a resolution better than $\sim$1\% and, therefore, 
``Centauro'' and/or strangelets events with an unusual ratio of electromagnetic to total 
(hadronic) energies~\cite{Gladysz-Dziadus:2001cq} can be well identified.\\

\begin{figure}[H] 
\begin{center}
\includegraphics[width=12cm]{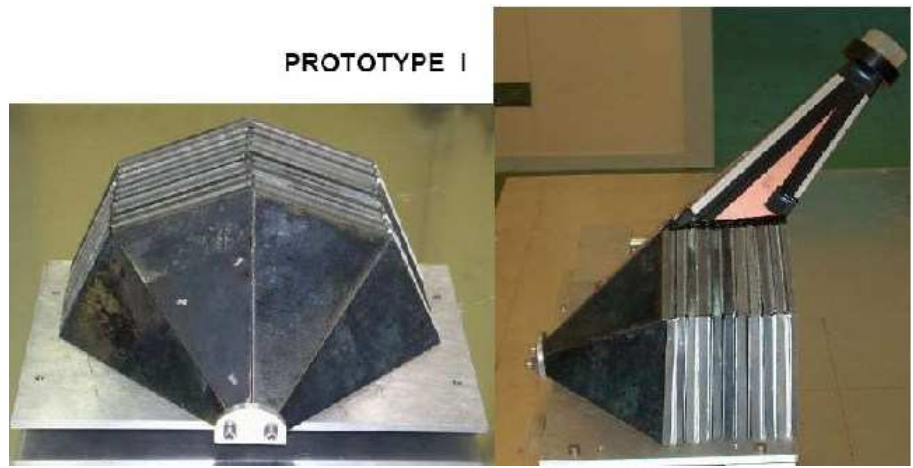}
\caption{CASTOR prototype I: frontal view (left picture) and lateral view (right picture, 
only one light guide is shown).}
\label{fig:castor}
\end{center}
\end{figure}

\begin{figure}[H] 
\begin{center}
\includegraphics[width=12cm,height=6cm]{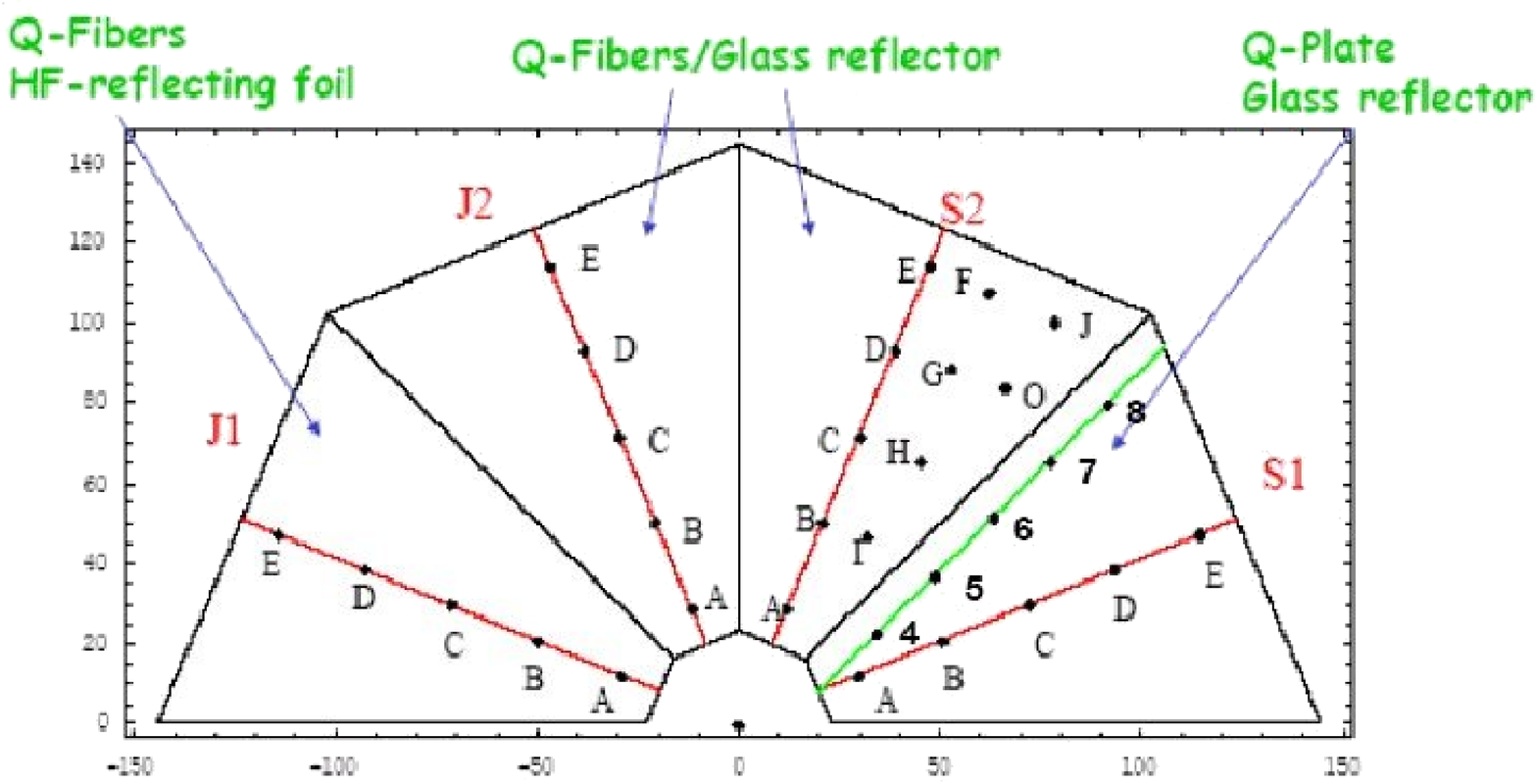}
\caption{Configuration options investigated in the 2003 beam test: 
different quartz structures (fibres and plate) and reflectors (glass, foil).
The points A-O and 4-8 are scan locations used in calorimeter response uniformity 
studies (see Section~\protect\ref{sec:area_scan}). $x-y$ units are mm.}
\label{fig:proto_scheme}
\end{center}
\end{figure}

\noindent
A calorimeter prototype has been constructed and tested with electron beams at CERN/SPS 
in the summer 2003. The purpose of this beam test was to investigate and compare the 
performance of different component options (structure of the quartz active material, 
choice of the light guides/reflectors and photodetector devices), rather than to obtain
precise quantitative results of the response of the final detector setup. The general 
view of the prototype is shown in Figure~\ref{fig:castor}. The different detector 
configurations considered in this work are shown schematically in Figure~\ref{fig:proto_scheme}. 
Preliminary results of the analysis have been presented at different CMS meetings~\cite{castor_meetgs}. 
Here we present a more quantitative analysis, including the beam profile data.

\section{Technical description}

The CASTOR detector is a \v{C}erenkov-effect based calorimeter with tungsten (W) 
absorber and quartz (Q) as sensitive material. An incident high-energy particle 
will shower in the tungsten volume and produce relativistic charged particles 
that will emit \v{C}erenkov light in the quartz plane. The \v{C}erenkov light is then 
collected and transmitted to photodetector devices through air-core light-guides. 
The different instrumentation options, investigated in this work, are shown in 
Figure~\ref{fig:proto_scheme}. In section~\ref{sec:WQ} we describe the various 
arrangements of the active (quartz) and passive (tungsten) materials of the 
calorimeter considered. Section~\ref{sec:light_guides} discusses the light 
transmission efficiency of different light-guide geometries, 
section~\ref{sec:light_guide_reflectors} compares two different light-guide 
reflecting materials, and section~\ref{sec:light_read} summarizes the characteristics 
of the photodetectors (photomultipliers and avalanche photodiodes) tested.


\subsection{Tungsten - Quartz}
\label{sec:WQ}

The calorimeter prototype is azimuthally divided into 4 octants and longitudinally 
segmented into 10 W/Q layers (Fig.~\ref{fig:castor}). Each tungsten absorber layer is 
followed by a number of quartz planes. The tungsten/quartz planes are inclined at 45$^\circ$ 
with respect to the beam axis to maximize \v{C}erenkov light output\footnote{The index 
of refraction of quartz is $n = 1.46-1.55$ for wavelengths $\lambda$ = 600-200 nm. The 
corresponding \v{C}erenkov threshold velocity is $\beta_c = 1/n = 0.65-0.69$, and therefore, 
for $\beta_c \approx$ 1 the angle of emission is $\theta_c = acos(1/n\beta) = 46^\circ-50^\circ$.}. 
The effective length of each W-plate is 7.07 mm, being inclined at 45$^\circ$. The total length 
is calculated to be 0.73$\lambda_{int}$ and 19.86$X_0$, taking a density for the used 
W-plates of $\sim$19.0 g/cm$^3$ and ignoring the contribution of the quartz material.

The calorimeter response and relative energy resolution were studied for quartz fibres 
(Q-F) and quartz plates (Q-P) (see Section~\ref{sec:beamtests}). We have tested four 
octant readout units of the calorimeter, arranged side-by-side in four azimuthal sectors. 
Each readout unit consisted of 10 sampling units.  Each sampling unit for sectors J1, J2, 
and S2 (see Fig.~\ref{fig:proto_scheme}) is comprised of a 5 mm thick tungsten plate
and three planes of 640 $\mu$m thick quartz fibres. The quartz fibres were produced
by Ceram Optec and have 600 $\mu$m pure fused silica core with a 40 $\mu$m polymer
cladding and a corresponding numerical aperture NA = 0.37 (in general, an optical fibre 
consists of the core with index of refraction $n_{core}$, and the cladding with index $n_{clad}$, 
and NA = $\sqrt{n^2_{core}-n^2_{clad}}$). The sampling unit
for sector S1 consisted of a 5 mm thick tungsten plate and one 1.8 mm thick quartz 
plate. Both types of quartz active material, fibre or plate, had about the same 
effective thickness. The filling ratio was 30\% and 37\% for the quartz fibres 
and quartz plates, respectively.


\subsection{Air-core light guides}
\label{sec:light_guides}

\begin{figure}[H]
\begin{center}
\resizebox{10cm}{!}
{\includegraphics{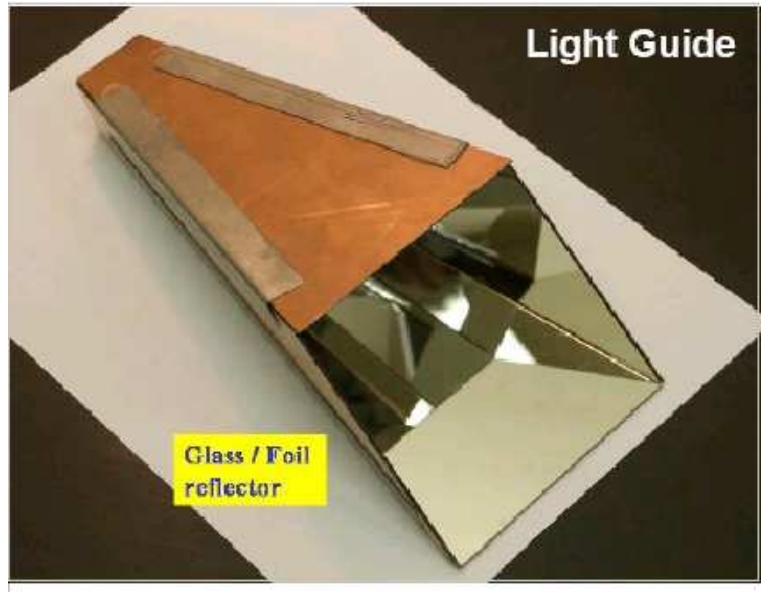}}
\caption{Picture of the light guide used in the prototype.}
\label{fig:light_guide}
\end{center}
\end{figure}

The light guide constructed for the CASTOR prototype I is shown in Figure~\ref{fig:light_guide}. 
It is an air-core light-guide made of Cu-plated 0.8 mm PVC (the internal walls are covered 
either with a glass reflector or with a reflector foil, which are compared in the next section). 
In this section the optimal design and dimensions of the light guide are obtained based on 
detailed {\sc geant} Monte Carlo simulations.

In the simulations, the \v{C}erenkov photons produced in the quartz of the calorimeter are 
collected and transmitted to the photodetectors by air-core light guides. The efficiency 
of light transmission and its dependence on the light-source position are crucial parameters 
characterizing the light guide and significantly affecting the performance of the calorimeter. 
We developed a {\sc geant} 3.21-based code to simulate the transmission of \v{C}erenkov photons
produced in the quartz plane through a light guide~\cite{mavro}. A photon is tracked until 
it is either absorbed by the walls or by the medium and is thus lost, or until it escapes 
from the light guide volume. In the latter case it is considered detected only if it escapes 
through the exit to the photodetector. If it is back-scattered towards the entry of 
the light guide it is also lost.

\begin{figure}[H] 
\begin{center}
\resizebox{10cm}{!}{
\includegraphics{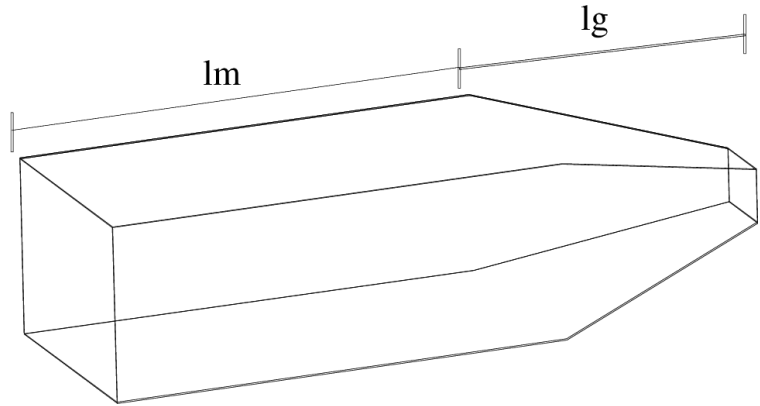}}
\caption{Schematic view of the air-core light guide geometry. $lg$ ($lm$)
is defined as the ratio of the length of the (non-)tapered section
over the width of the entrance plane (assumed to be unity in the figure).}
\label{fig:light_guide_geom}
\end{center}
\end{figure}

Inside the fibre core \v{C}erenkov photons are practically produced isotropically. But those
that are captured and propagate through the lightguide have an exit angle with
respect to the fibre longitudinal axis up to a maximum value ($\theta_{core}$) which
depends on the numerical aperture NA and the core refraction index ($n_{core}$). When
traversing the core-air boundary at the entrance of the lightguide, the photons undergo
refraction resulting in a larger angle ($\theta_{air}$).
In the simulations, fibres of various numerical apertures (NA = 0.22 - 0.48) as well as 
light-guides of various shapes (fully square cross section or partially tapered) 
were used (see Fig.~\ref{fig:light_guide_geom}).  
The maximum values of core-exiting and air-entering angles ($\theta_{core}$, $\theta_{air}$) 
in degrees for various numerical apertures are given in Table~\ref{tab:1}. For the quartz 
plate, the air-entering angle, $\theta_{air}$, is larger than 30$^\circ$. 

\begin{table}[H]
\begin{center}
\caption{Maximum values of the core-exiting ($\theta_{core}$) and air-exiting ($\theta_{air}$) 
angles, for various numerical apertures (NA) of the quartz fibres (index of refraction: $n_{core}$ = 1.46).}
\vskip 0.2cm
\label{tab:1}
\renewcommand{\arraystretch}{1.5}
\begin{tabular}{ccc} \hline
   NA ($n_{core}$=1.46) & $\theta_{core}$ & $\theta_{air}$\\ \hline  
      0.22 	  & 8.7 	& 	12.7 \\  
      0.37 	& 	14.7 	& 	21.7 \\   
      0.40 	& 	15.9 	& 	23.6 \\  
      0.44 	& 	17.5 	& 	26.1 \\  
      0.48 	& 	19.2    & 	28.7 \\  \hline 
\end{tabular}
\renewcommand{\arraystretch}{1}
\end{center}
\end{table}

\begin{figure}[H] 
\begin{center}
\resizebox{10cm}{!}
{\includegraphics{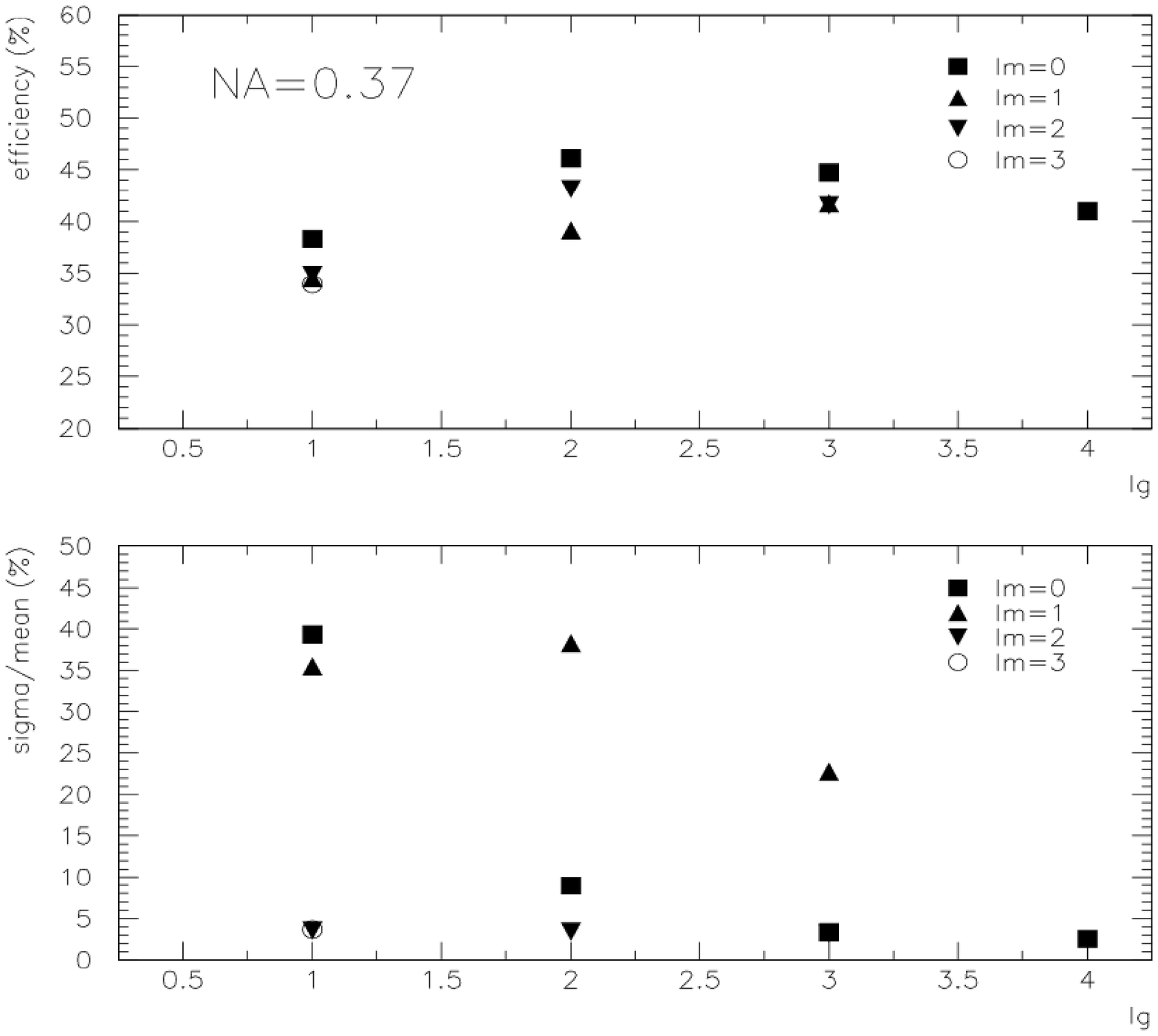}}
\caption{Efficiency (top) and relative variation of the efficiency (bottom) 
for various light guides (calorimeter quartz fibres with NA = 0.37) 
for different values of the $lg$ and $lm$ parameters (see text).}
\label{fig:effic_lightguide_0.37}
\end{center}
\end{figure}

\begin{figure}[H] 
\begin{center}
\resizebox{10cm}{!}
{\includegraphics{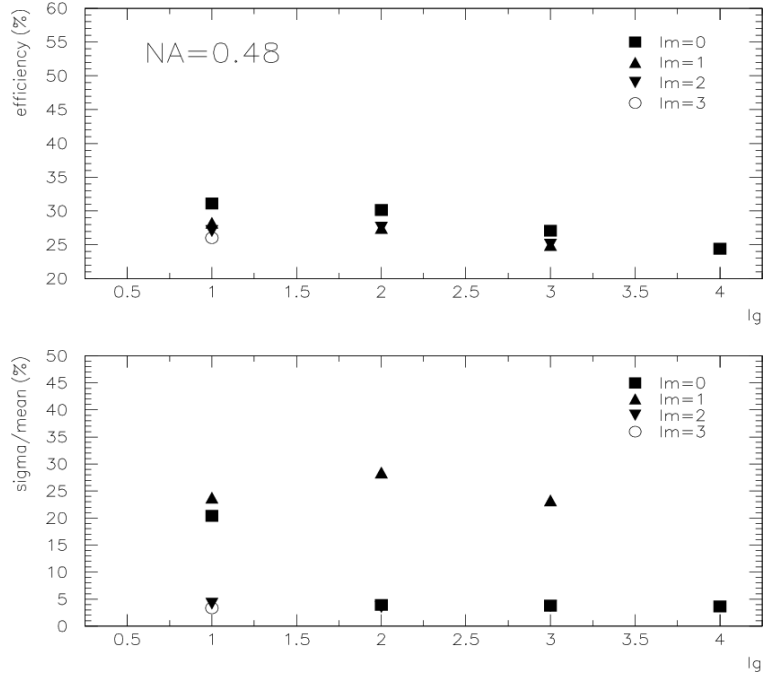}}
\caption{Efficiency (top) and relative variation of the efficiency (bottom) 
for various light guides (calorimeter quartz fibres with NA = 0.48) 
for different values of the $lg$ and $lm$ parameters (see text).}
\label{fig:effic_lightguide_0.48}
\end{center}
\end{figure}

The walls of the {\sc geant} light-guide have a reflection coefficient of 0.85 (simulating 
the transmittance of the reflecting internal mirror surface and the quantum efficiency
of the photodetector devices, see next Section and Table~\ref{tab:transmittance}). 
The entrance plane of the light guide was uniformly scanned with the simulated light source. 
The percentage of photons escaping in the direction of the photodetector has been recorded 
as a function of the source position, giving, after integration over the complete surface, the light guide efficiency. 
The spatial uniformity of the light-guide performance can be quantified with the relative 
variation ($\sigma/$mean) of the efficiency across the entrance. Results for the light guides 
efficiency and uniformity studied are tabulated\footnote{Note, that only the points relevant for 
the actual light-guide construction are included in the table.} in Tables~\ref{tab:1}--\ref{tab:5} 
and are plotted in Figures~\ref{fig:effic_lightguide_0.37} and ~\ref{fig:effic_lightguide_0.48} 
for fibres with NA = 0.37 and 0.48, respectively. We studied air-core lightguides of square
cross section (with entrance area 10$\times$10 cm$^2$), fully or partially tapered.
The parameters $lg$ and $lm$ refer to the tapered and non-tapered sections of the light guide, 
as shown in Figure~\ref{fig:light_guide_geom}, defined as~\cite{mavro}:
\begin{description}
\item $lg$ = ratio of the length of the tapered part over the width of the entrance plane, and
\item $lm$ = ratio of the length of non tapered part over the width of the entrance plane.
\end{description}
Thus, e.g. with a mean entrance length of 10 cm, a value $lg:lm$=1:2 indicates that the 
light-guide has a total length of 30 cm with 10 cm of tapering part, and a value 
$lg:lm$=2:0 indicates a fully tapered light-guide with length 20 cm, and so on.
In tables~\ref{tab:2}--\ref{tab:5}, the row (column) indicates the magnitude of the 
parameters $lm$ ($lg$), respectively. 

\begin{table}[H]
\begin{center}
\caption{Light-guide efficiency (\%) for different values of the $lg$ 
and $lm$ parameters (see text) and quartz fibres with NA = 0.37.}
\vskip0.2cm
\label{tab:2}
\renewcommand{\arraystretch}{1.5}
\begin{tabular}{lccc} \hline
$_{\textstyle lg}\backslash^{\textstyle lm}$ & 0 &	1 	&	2    \\\hline
1 &	38.3 	&	34.5 	&	34.8 \\	
2 &	46.1 	&	39.1 	&	43.2 \\
3 &	44.8 	&	41.8 	&	41.5 \\\hline 
\end{tabular}
\renewcommand{\arraystretch}{1}
\end{center}
\end{table}

\begin{table}[H]
\begin{center}
\caption{ Relative variation of the light-guide efficiency across the 
entrance, $\sigma/$Mean (\%), for different values of the $lg$ and $lm$ 
parameters (see text) and quartz fibres with NA = 0.37.}
\vskip0.2cm
\label{tab:3}
\renewcommand{\arraystretch}{1.5}
\begin{tabular}{lccc} \hline
$_{\textstyle lg}\backslash^{\textstyle lm}$ & 0 &	1 	&	2   \\\hline
1 & 	39.3 & 		35.5  &		3.6 \\		
2 & 	8.9  &		38.3  &		3.4 \\	
3 & 	3.3  &		22.8  &		3.2 \\\hline 	
\end{tabular}
\renewcommand{\arraystretch}{1}
\end{center}
\end{table}

\begin{table}[H]
\begin{center}
\caption{ Light-guide efficiency (\%) for different values of the $lg$ 
and $lm$ parameters (see text) and quartz fibres with NA = 0.48.}
\vskip0.2cm
\label{tab:4}
\renewcommand{\arraystretch}{1.5}
\begin{tabular}{lccc} \hline
$_{\textstyle lg}\backslash^{\textstyle lm}$ & 0 &	1 	&	2    \\\hline
1  &	31.1  &		28.3  &		27.1 \\	 
2  &	30.1  &		27.5  &		27.5 \\
3  &	27.1  &		25.0  &		25.0 \\\hline 	
\end{tabular}
\renewcommand{\arraystretch}{1}
\end{center}
\end{table}

\begin{table}[H]
\begin{center}
\caption{Relative variation of the light-guide efficiency across the 
entrance, $\sigma/$Mean (\%), for different values of the $lg$ and $lm$ 
parameters (see text) and quartz fibres with NA =  0.48.}
\vskip0.2cm
\label{tab:5}
\renewcommand{\arraystretch}{1.5}
\begin{tabular}{lccc} \hline
$_{\textstyle lg}\backslash^{\textstyle lm}$ & 0 &	1 	&	2   \\\hline
1 &	20.4 &		23.8 	&	4.1 \\	 
2 &	3.9 &		28.4 	&	4.6 \\
3 &	3.8 &		23.2 	&	3.7 \\\hline 	
\end{tabular}
\renewcommand{\arraystretch}{1}
\end{center}
\end{table}

From the tables~\ref{tab:1}-\ref{tab:5} and figures~\ref{fig:effic_lightguide_0.37} and 
\ref{fig:effic_lightguide_0.48} we note that, as the NA of the fibre and hence the air-entering 
angle, $\theta_{air}$, increases, the transmission efficiency decreases. Also, the optimum length 
for the air-core light guide decreases, while the uniformity of the light exiting increases.
In order to obtain an optimum efficiency and uniformity of light transmission within the 
realistically available space, the best option seems $lm$ = 0 and $lg$ = 2 for NA = 0.37 and 0.48.
A more detailed study of the light guide performances -- beyond the scope of our current paper -- 
can be found in reference~\cite{mavro}. 


\subsection{Light guide reflecting material}
\label{sec:light_guide_reflectors}

\begin{figure}[H] 
\begin{center}
\includegraphics[width=14cm]{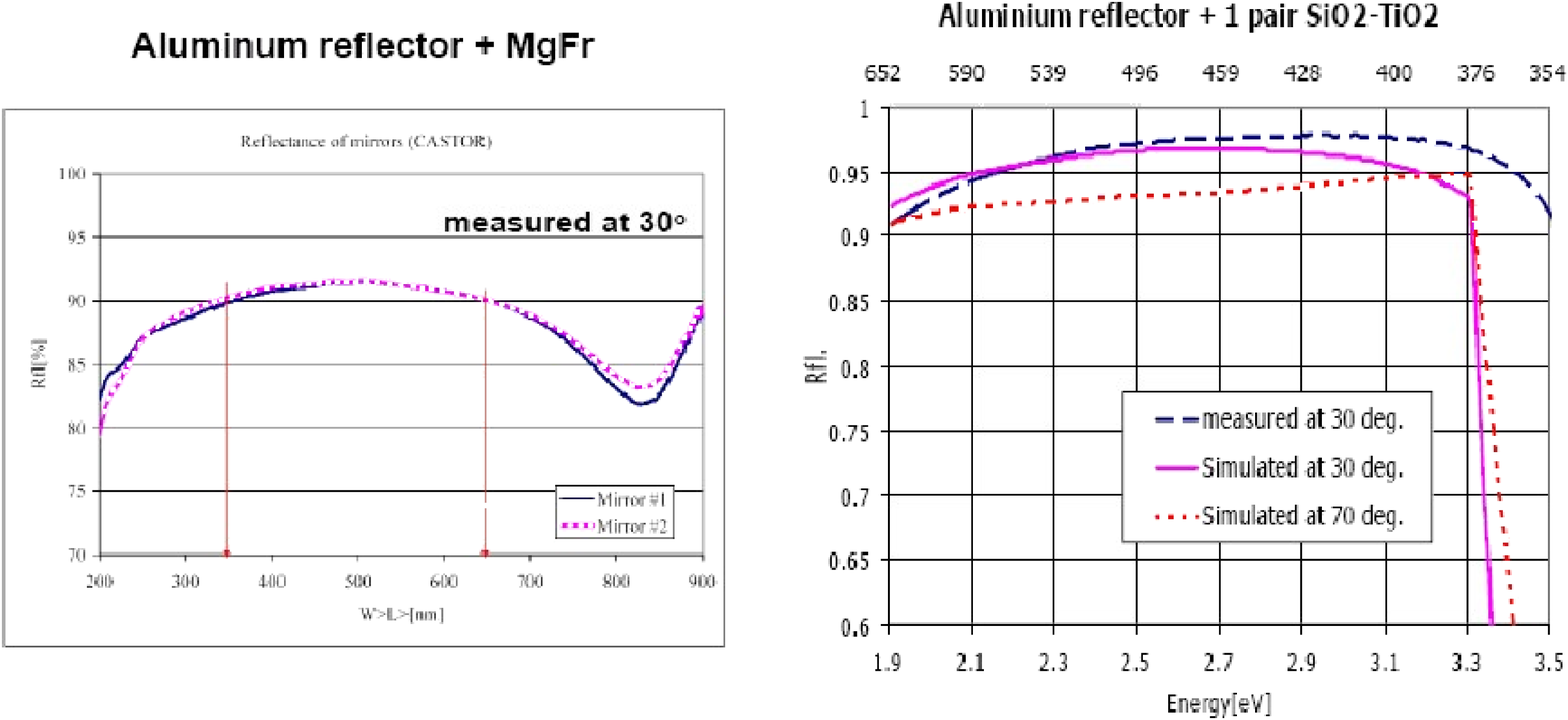}
\caption{Reflectance of two mirrors coated with (a) AlO+MgFr and (b) Dupont foil 
with AlO and SiO$_2$+TiO$_2$, as a function of the incident light wavelength.}
\label{fig:reflectance_vs_wavelength}
\end{center}
\end{figure}

The light transmittance in the light-guides was studied for two alternatives for the reflecting 
medium:
\begin{enumerate}
\item 0.5 mm thick float-glass with evaporations of AlO and MgFr (Fig.~\ref{fig:reflectance_vs_wavelength}a) 
and 
\item Dupont polyester film reflector coated with AlO and reflection enhancing 
dielectric layer stack SiO$_2$+TiO$_2$, the so-called HF reflector foil
(Fig.~\ref{fig:reflectance_vs_wavelength}b).
\end{enumerate}

To choose the most suitable reflector, we also have to take into account the quantum 
efficiency of the photodetector device (see Section~\ref{sec:light_read}). 
In Table~\ref{tab:transmittance} we calculate the product of the light guide transmittance 
and Avalanche Photodiodes (APD) quantum efficiency for Q-fibres with NA = 0.37 
and 3 internal reflections in the designed light guide. The light output is higher (lower) 
for the light-guides with reflector-foil (glass-reflector) for wavelengths above (below) 
$\lambda$ = 400 nm. We prefer the HF-reflector solution since the short wavelength 
\v{C}erenkov light ($\lambda$ $<$ 400 nm) deteriorates fast with irradiation of the 
quartz material and thus a continuous compensation must be applied. The optimum 
combination of the HF-reflector and the Q-efficiency of the photodetector ensures 
that the total efficiency is maximized above 400 nm and falls sharply to zero below 400 nm.

\begin{table}[H]
\begin{center}
\caption{Light guide transmittance times the Avalanche Photodiode quantum efficiency 
at each wavelength (see Figure~\ref{fig:quantum_effic}) for the two reflectors considered 
(in both cases the quartz fibres have NA = 0.37 and 3 internal reflections).}
\vskip0.2cm
\label{tab:transmittance}
\renewcommand{\arraystretch}{1.5}
\begin{tabular}{ccc} \hline
Wavelength  &   Glass reflector (Al+MgF) & Dupont + Layer stack \\\hline
650 nm &	62\% &	64\% \\	 
400 nm &	53\% &	62\% \\
350 nm &	44\% &  7\%\\ 	
300 nm &	10\% &  $\sim$0\%\\\hline 	
\end{tabular}
\renewcommand{\arraystretch}{1}
\end{center}
\end{table}


\subsection{Photodetectors}
\label{sec:light_read}

\begin{figure}[H] 
\begin{center}
\includegraphics[width=11cm,height=5cm]{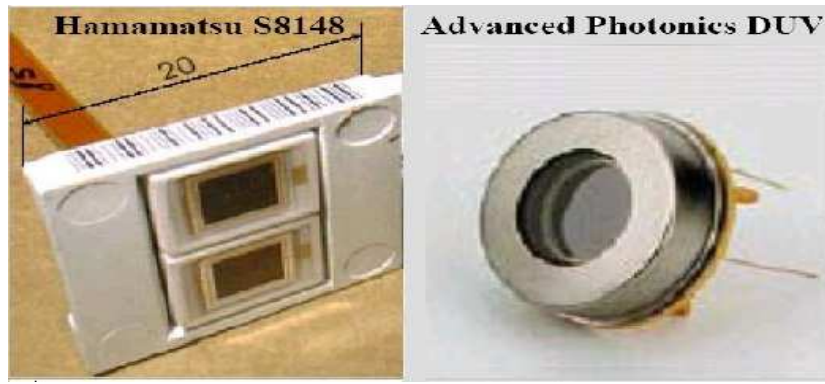}
\caption{The two types of APDs used in the beam test: Hamamatsu S8148
(left, 5$\times$5 mm$^2$, in a 2$\times$2 matrix with total 1 cm$^2$ active area) 
and Advanced Photonix DUV (right, active area of 2 cm$^2$).}
\label{fig:apds}
\end{center}
\end{figure}

We instrumented the calorimeter prototype with two different types of light-sensing devices:
\begin{enumerate}
\item Two different kinds of Avalanche Photodiodes (APDs): Hamamatsu S8148 (APD1, 
developed for the CMS electromagnetic calorimeter~\cite{apd1}) and 
Advanced Photonix Deep-UV (APD2), Fig.~\ref{fig:apds}.
\item Two different types of photomultipliers (PMTs): Hamamatsu R374 and Philips XP2978.
\end{enumerate}

\begin{figure}[H] 
\begin{center}
\includegraphics[width=6.5cm]{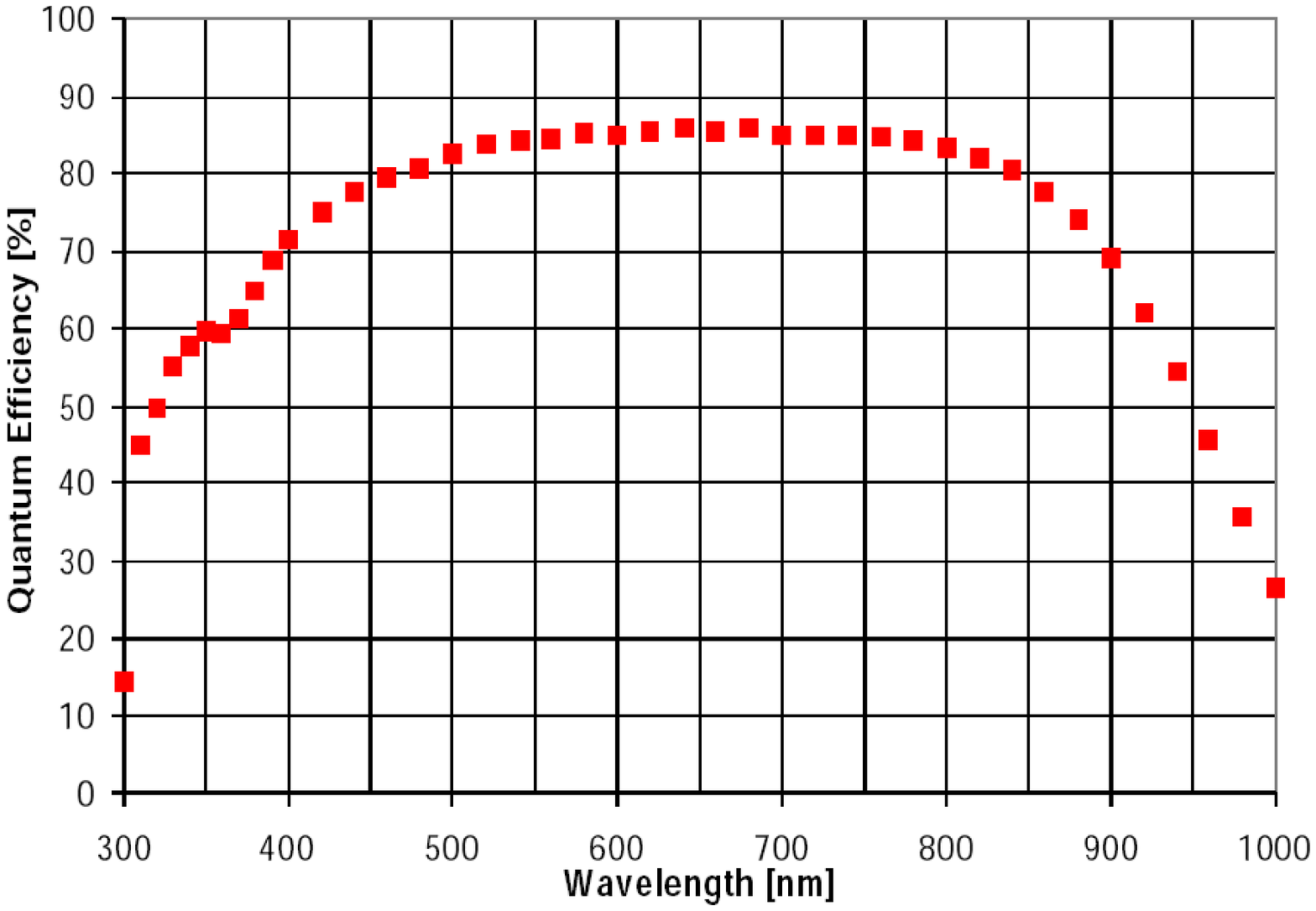}
\includegraphics[width=7cm]{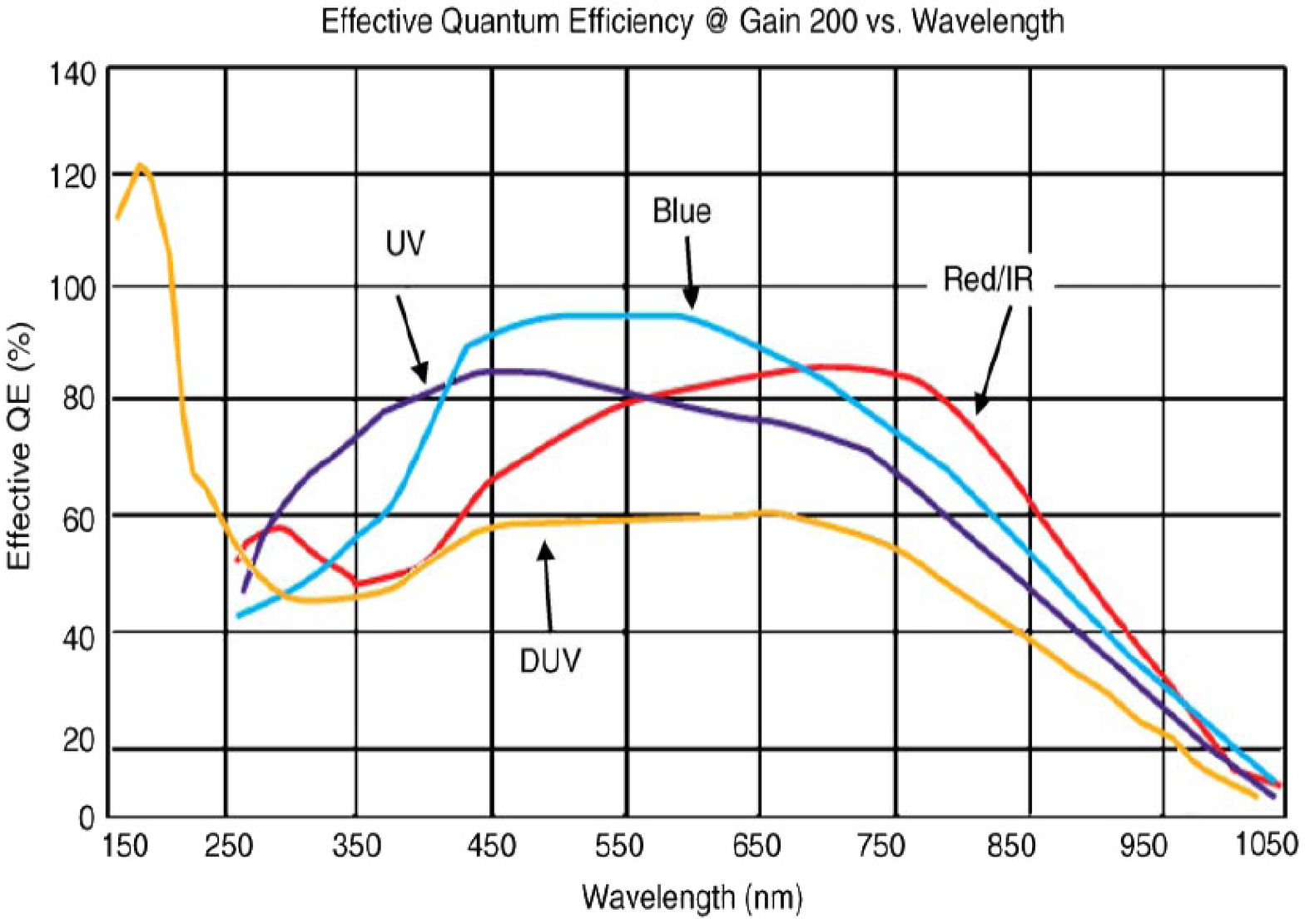}
\caption{APDs quantum efficiencies versus wavelength: Hamamatsu S8148 (left) 
and Advanced Photonix (right, the curve labeled 'blue' is relevant for this study).}
\label{fig:quantum_effic}
\end{center}
\end{figure}

We used 4 Hamamatsu APDs, each 5$\times$5 mm$^2$, in a 2$\times$2 matrix with total area 
of 1 cm$^2$. The Advanced Photonix DUV APD had an active area of 2 cm$^2$ (16 mm diameter). 
The Hamamatsu and Philips PMTs have both an active area of 3.1 cm$^2$. The Hamamatsu 
and Advanced Photonix APD quantum efficiencies are shown versus wavelength in 
Fig.~\ref{fig:quantum_effic}.

\section{Beam Test Results}
\label{sec:beamtests}

The beam test took place in summer 2003 at the H4 beam line of the CERN SPS. The 
calorimeter prototype was placed on a platform movable with respect to the electron 
beam in both horizontal and vertical (X,Y) directions. Telescopes of two wire chambers, 
as well as two crossed finger scintillator counters, positioned in front of the calorimeter, 
were used to determine the electron impact point. In the next two sections we present the 
measured calorimeter linearity and resolution as a function of energy and impact point 
for different prototype configurations.

\subsection{Energy Linearity and Resolution}

\begin{figure}[H] 
\begin{center}
\resizebox{12cm}{!}
{\includegraphics[angle=-90]{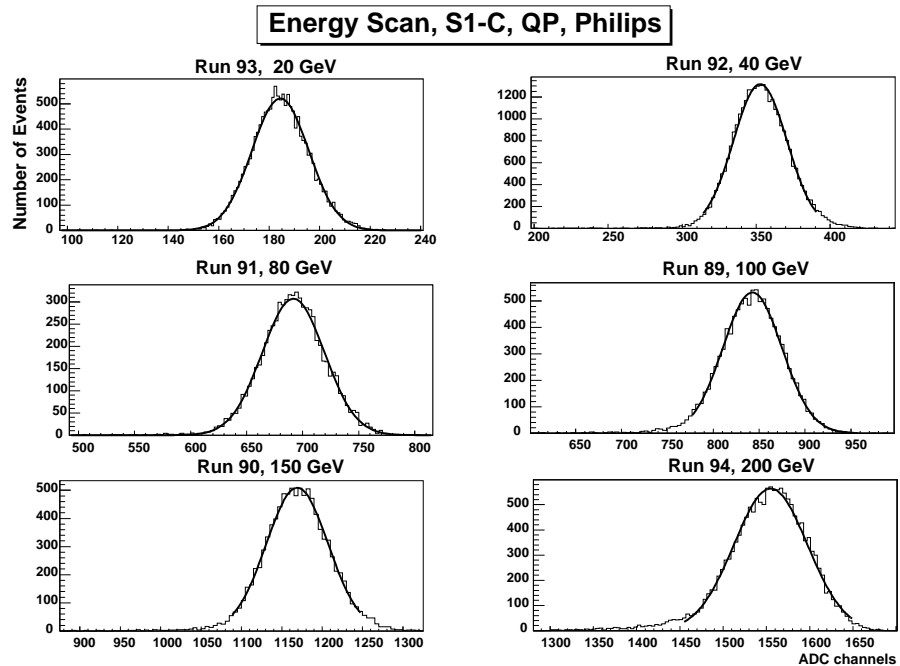}}
\caption{Distributions of signal amplitudes in ADC channels for 
electron beam energies (20, 40, 80, 100, 150 and 200 GeV) 
impinging on the central point C of sector S1 (Quartz-Plate) using 
Philips PMT.}
\label{fig:adc_signals1}
\end{center}
\end{figure}

\begin{figure}[H] 
\begin{center}
\resizebox{12cm}{!}
{\includegraphics[angle=-90]{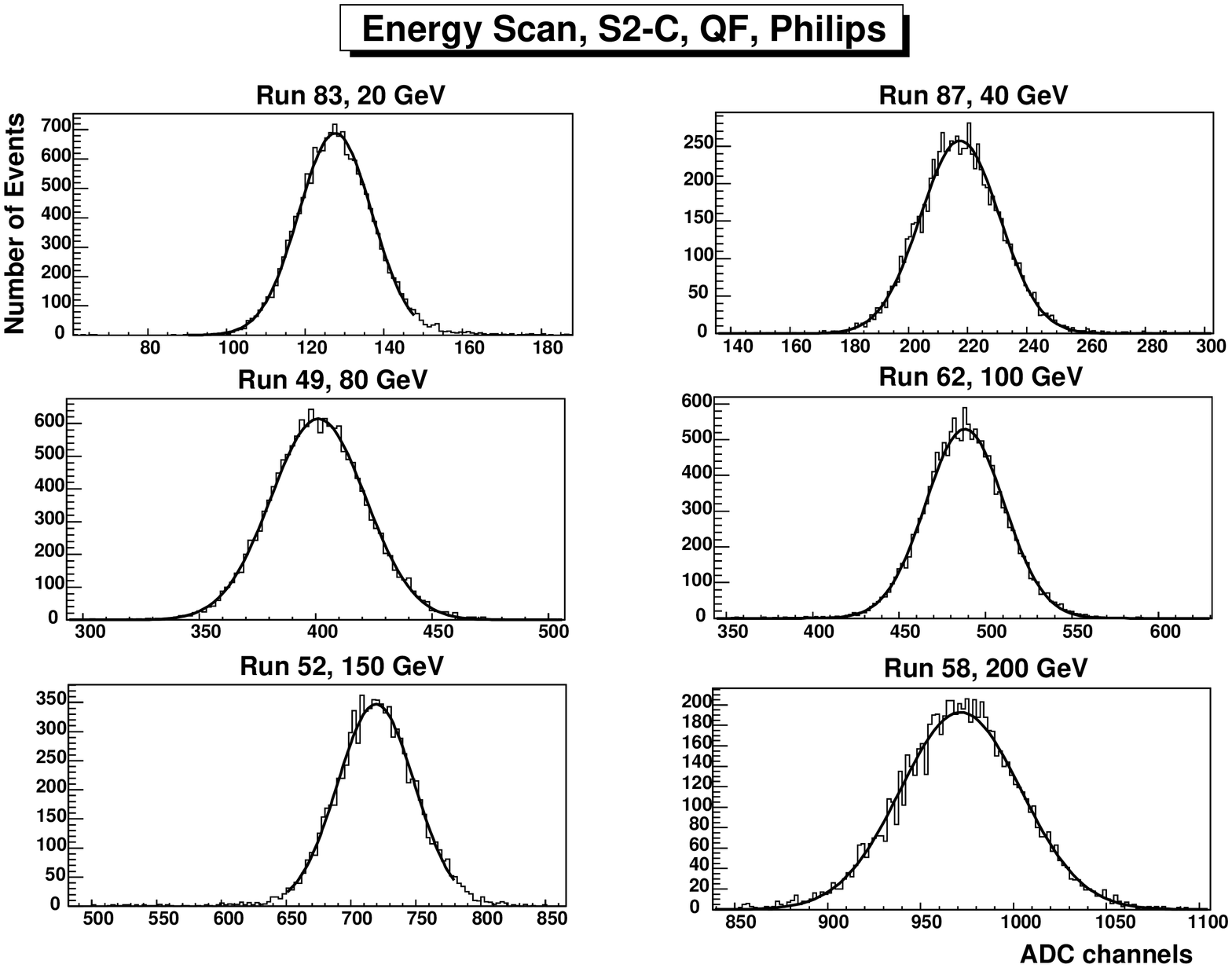}}
\caption{Distributions of signal amplitudes in ADC channels for 
electron beam energies (20, 40, 80, 100, 150 and 200 GeV) 
impinging on the central point C of sector S2 (Quartz-Fibre) using Philips PMT.}
\label{fig:adc_signals2}
\end{center}
\end{figure}

\begin{figure}[H] 
\begin{center}
\resizebox{12cm}{!}
{\includegraphics[angle=-90]{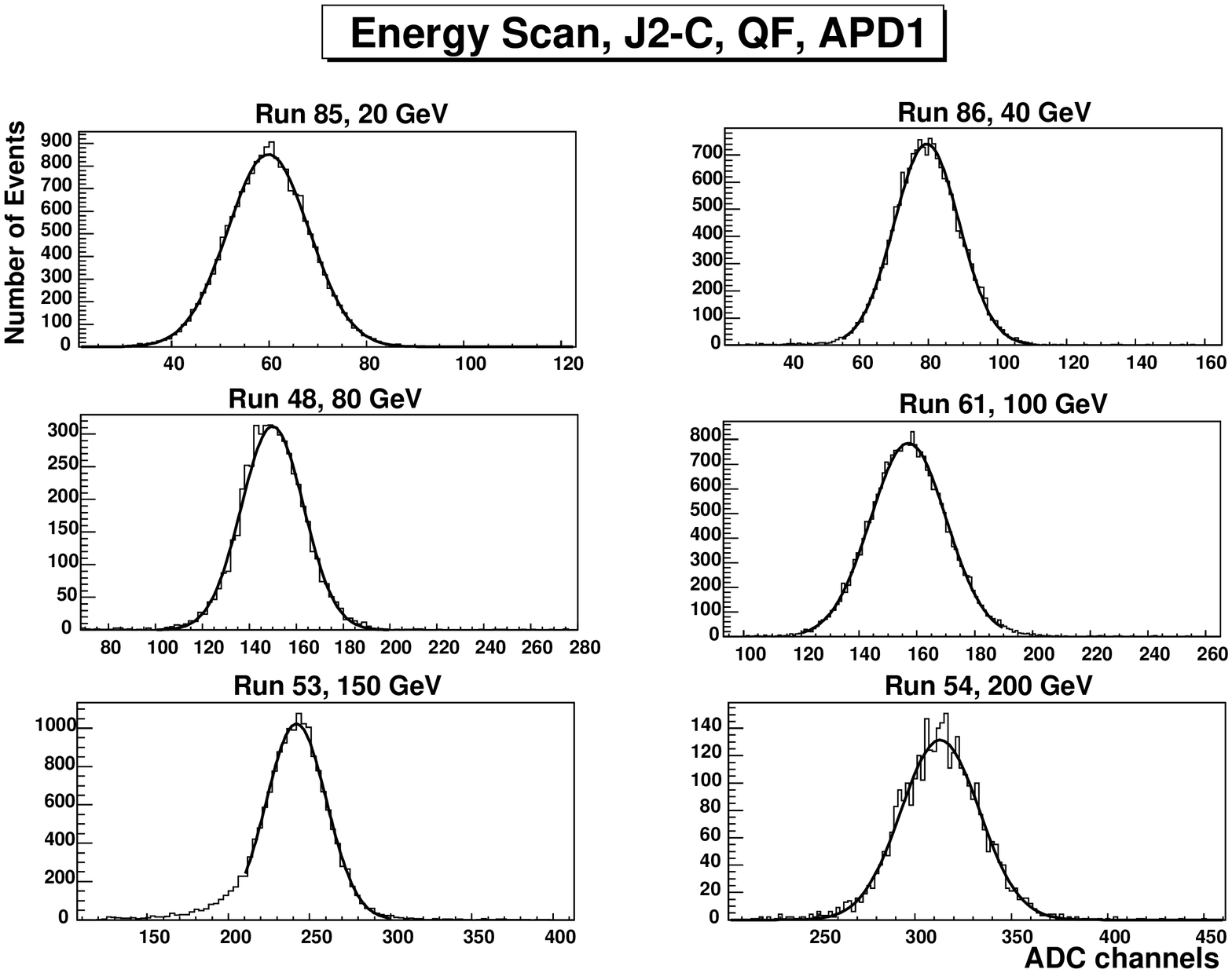}}
\caption{Distributions of signal amplitudes in ADC channels for 
electron beam energies (20, 40, 80, 100, 150 and 200 GeV) 
impinging on the central point C of sector J2 (Quartz-Fibre) using Hamamatsu APD.}
\label{fig:adc_signals3}
\end{center}
\end{figure}

\begin{figure}[H] 
\begin{center}
\resizebox{12cm}{!}
{\includegraphics[angle=-90]{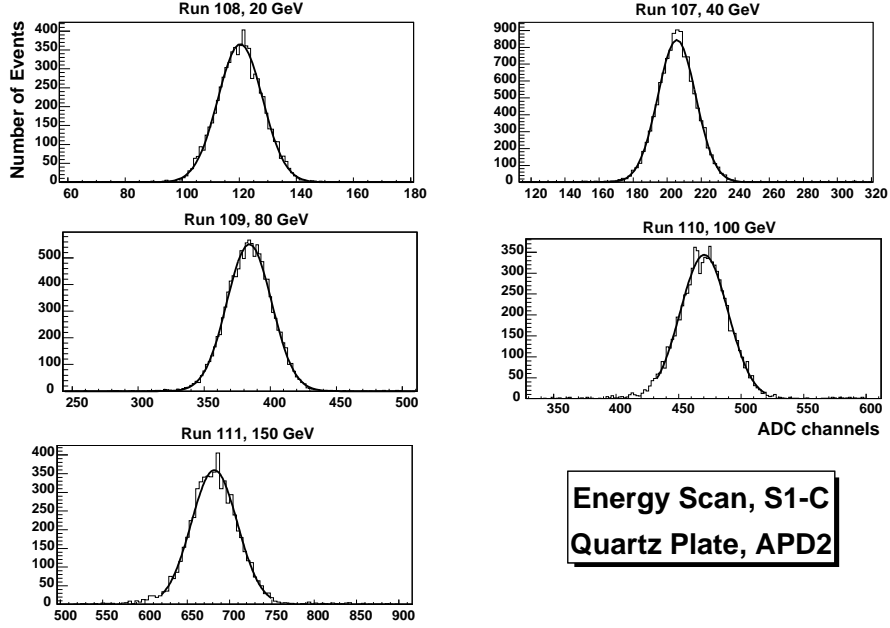}}
\caption{ Distributions of signal amplitudes in ADC channels for 
electron beam energies (20, 40, 80, 100, and 150 GeV) 
impinging on the central point C of sector S1 (Quartz-Plate) using Advanced Photonix APD.}
\label{fig:adc_signals4}
\end{center}
\end{figure}

To study the linearity of the calorimeter response and the relative energy resolution as 
a function of energy, the central points C (Fig.~\ref{fig:proto_scheme}) in different 
azimuthal sectors have been exposed to electron beams of energy 20, 40, 80, 100, 150 
and 200 GeV. The results of the energy scanning, analyzed for four calorimeter configurations, 
are shown in figures~\ref{fig:adc_signals1}--\ref{fig:adc_signals4}. 
The distributions of signal amplitudes, after introducing the cuts accounting for the 
profile of the beam, are symmetric and well fitted by a Gaussian function.

\begin{figure}[H] 
\begin{center}
\includegraphics[width=5.3cm,angle=-90]{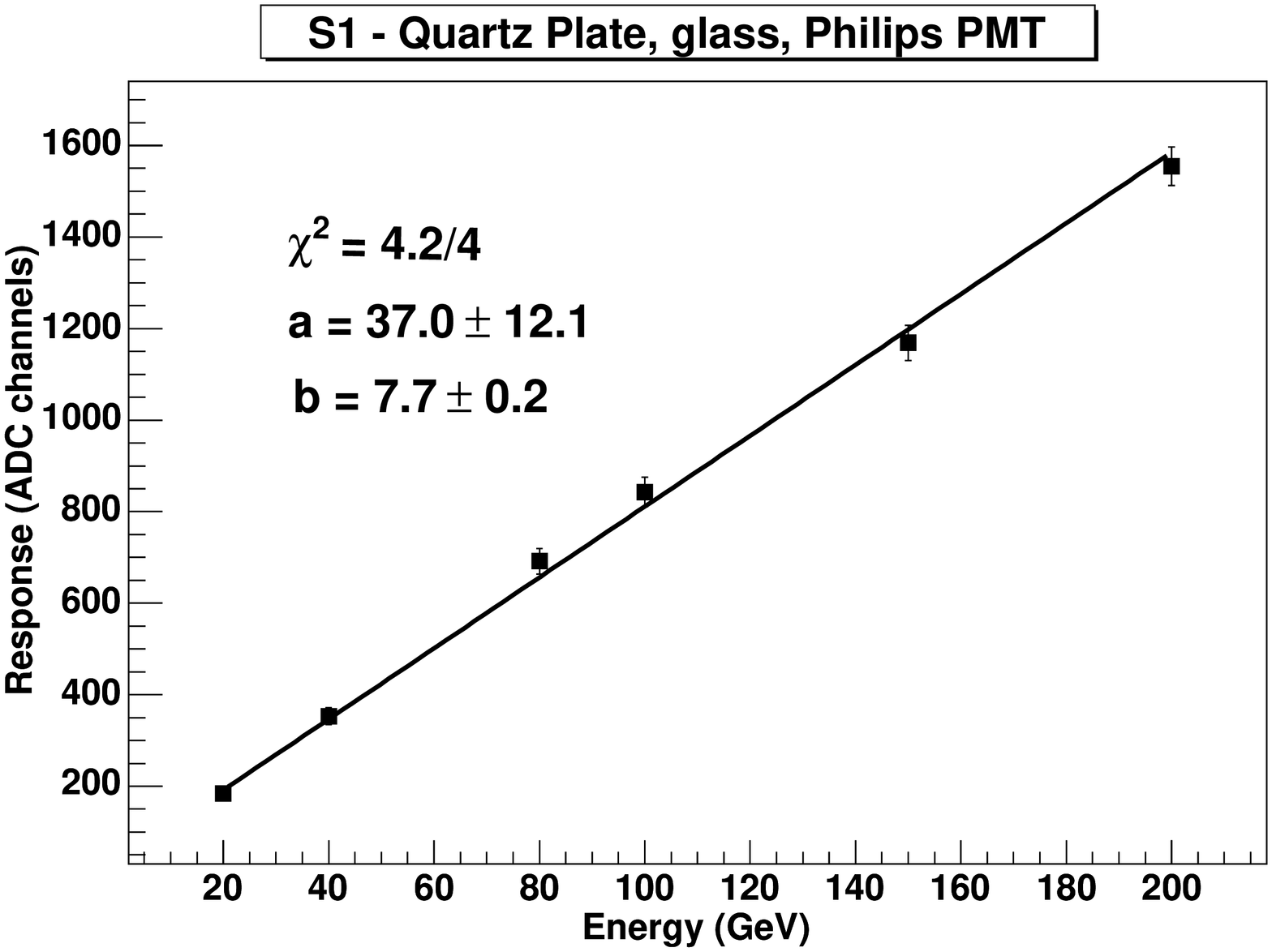}
\includegraphics[width=5.3cm,angle=-90]{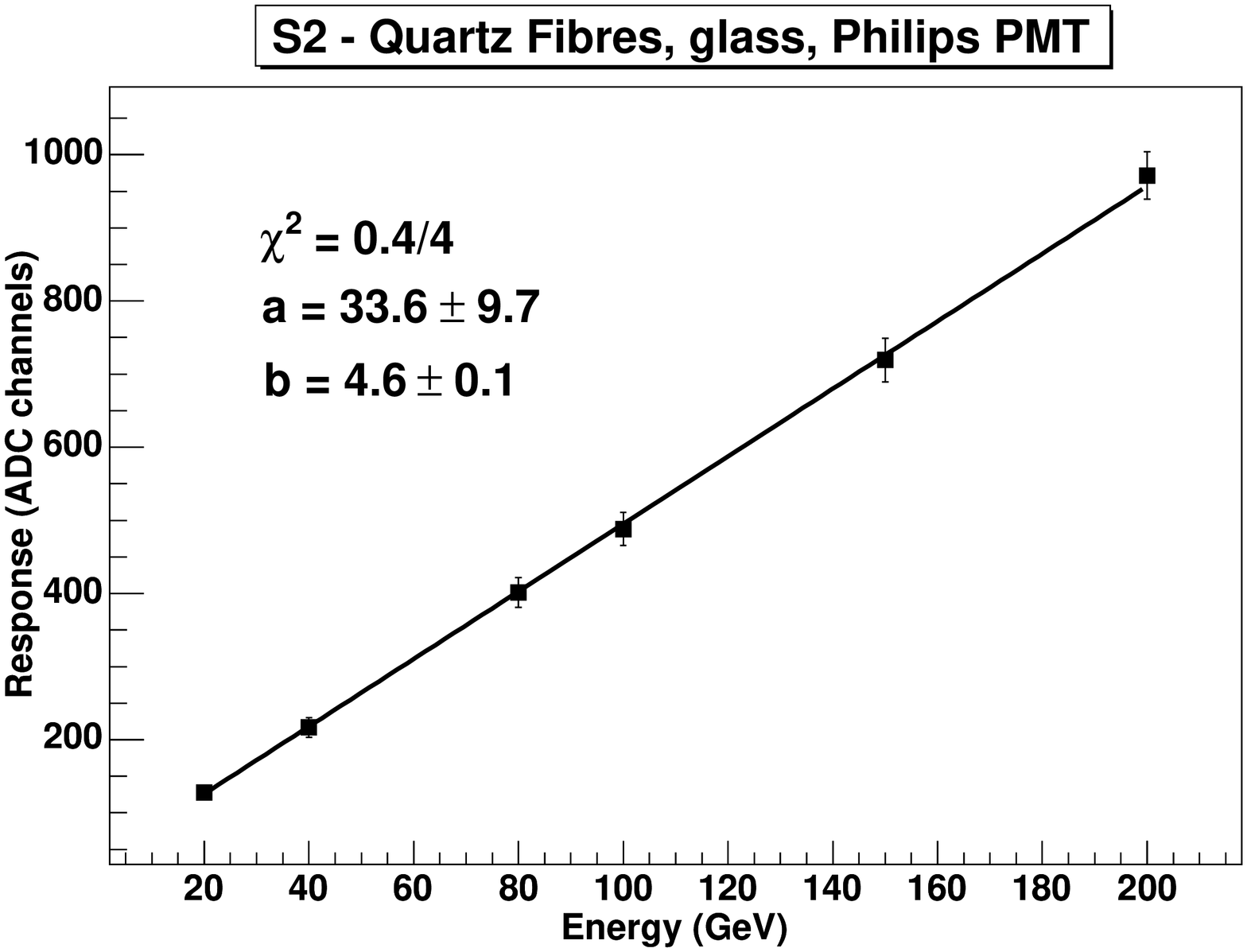}\vspace{1mm}
\includegraphics[width=5.3cm,angle=-90]{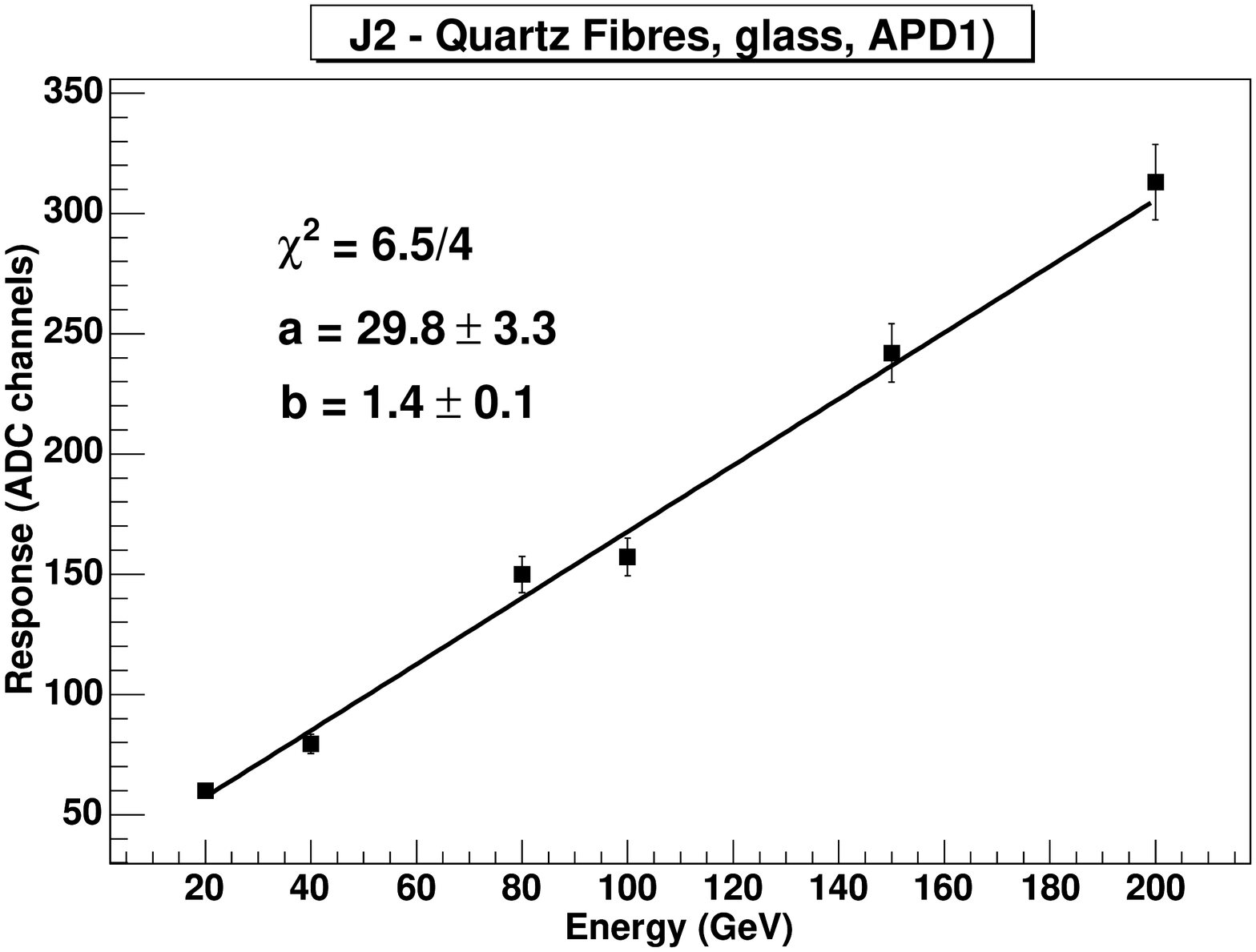}
\includegraphics[width=5.3cm,angle=-90]{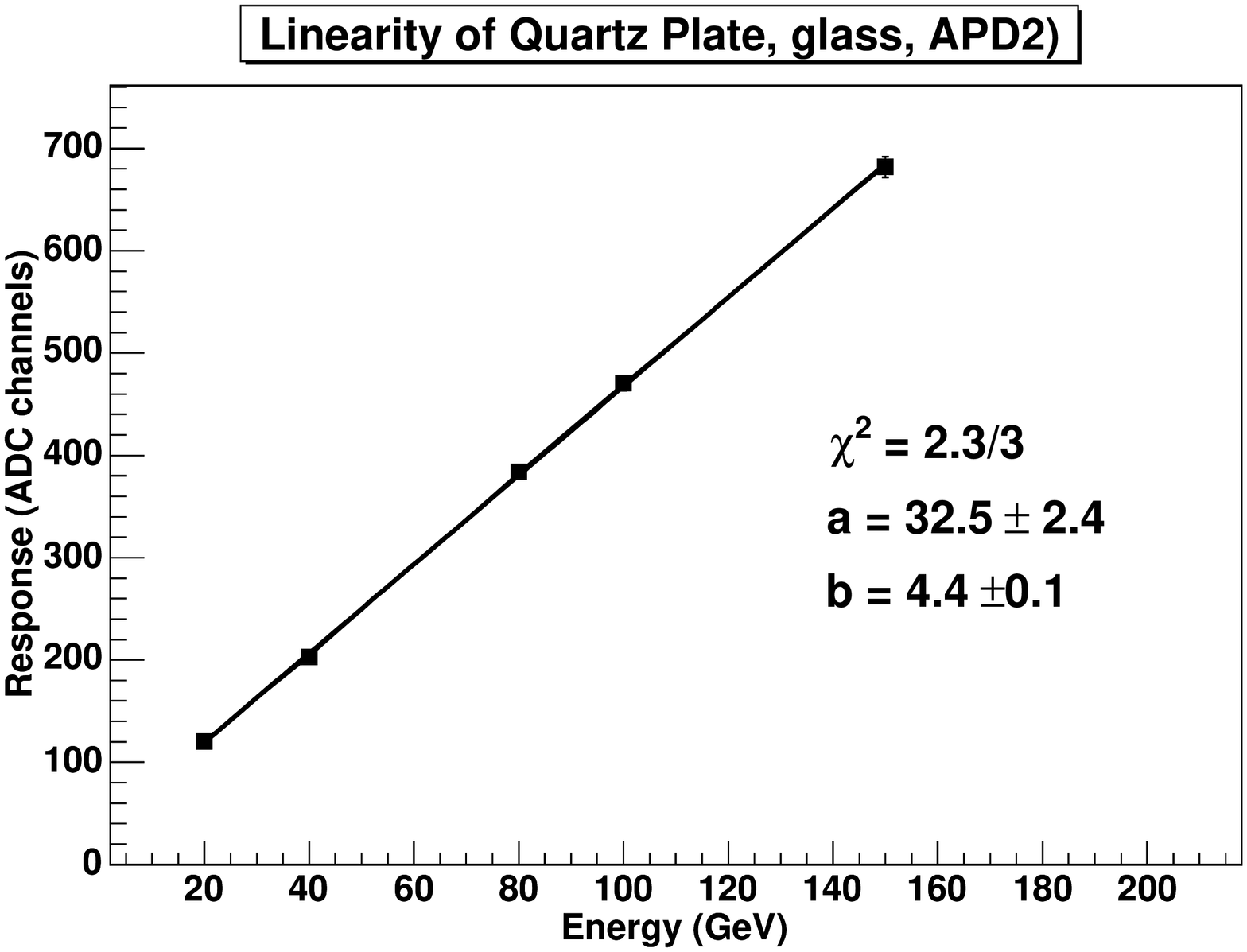}
\caption{Energy linearity in sectors: (a) S1 (Philips PMT), 
(b) S2 (Philips PMT), (c) J2 (APD1), (d) S1 (APD2).}
\label{fig:linearity}
\end{center}
\end{figure}

For all configurations, the calorimeter response is found to be linear in the energy 
range explored (see Fig.~\ref{fig:linearity}). The average signal amplitude, 
expressed in units of ADC channels, can be satisfactorily fitted by the following formula: 

\begin{eqnarray}
ADC & = & a + b \times E 
\end{eqnarray}

where the energy $E$ is in GeV. The fitted values of the parameters for 
each configuration are shown in Fig.~\ref{fig:linearity} and are tabulated in 
Table~\ref{tab:linearity_resol}. The values of the intercept 'a' are consistent
with the position of the ADC pedestal values measured for the various 
configurations considered: 36.1 $\pm$ 0.3 (S1-Quartz Plate), 
38.4 $\pm$ 1.8 (S2-Quartz Fibres), 35.3 $\pm$ 1.5 (J2-Quartz Fibres, 
glass reflector), 35.4 $\pm$ 0.6 (J1-Quartz Fibres, foil reflector).

The relative energy resolution of the calorimeter has been studied by plotting the 
normalized width of the Gaussian signal amplitudes (Figs.~\ref{fig:adc_signals1}--
\ref{fig:adc_signals4}), $\sigma/E$, with respect to the incident beam electron energy, 
E (GeV) and fitting the data points with two different functional forms~\cite{resolution}: 

\begin{eqnarray}
\sigma/E & = & p_0 + p_1/\sqrt{E} 			\label{eq:2} \\
\sigma/E & = & p_0 \oplus p_1/\sqrt{E} \oplus p_2/E 	\label{eq:3}
\end{eqnarray}
 
where the $\oplus$ indicates that the terms have been added in quadrature. 
In expression~(\ref{eq:3}), three terms determine the energy resolution:
\begin{enumerate}
\item The constant term $p_0$, coming from the gain variation with changing voltage and 
temperature, limits the resolution at high energies. 
\item The dominant stochastic term $p_1$, due to intrinsic shower photon statistics.
\item The noise $p_2$ term, which contains the noise contribution from capacitance and dark current. 
\end{enumerate}

\begin{figure}[H] 
\begin{center}
\includegraphics[width=6.8cm]{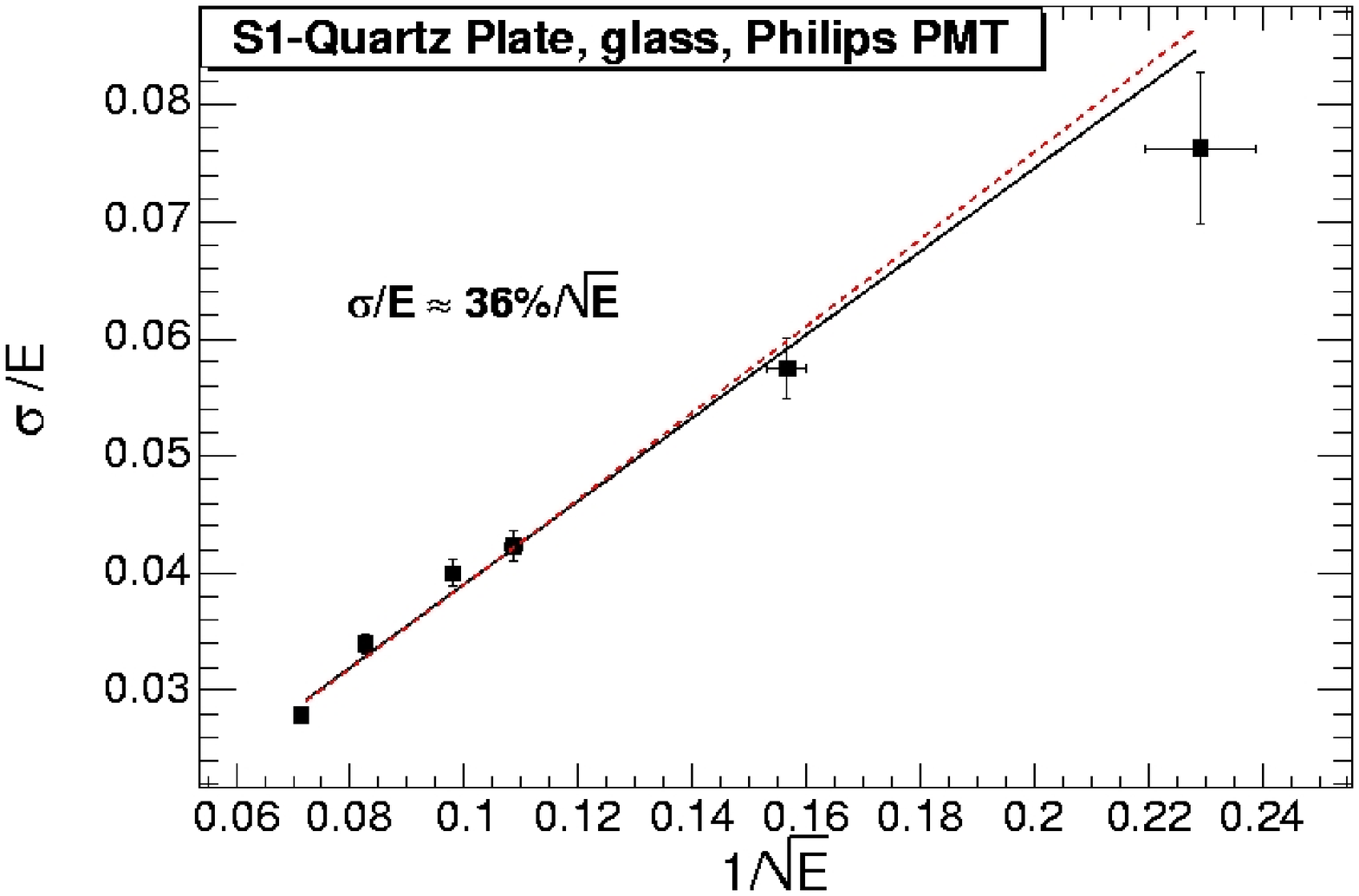}
\includegraphics[width=6.8cm]{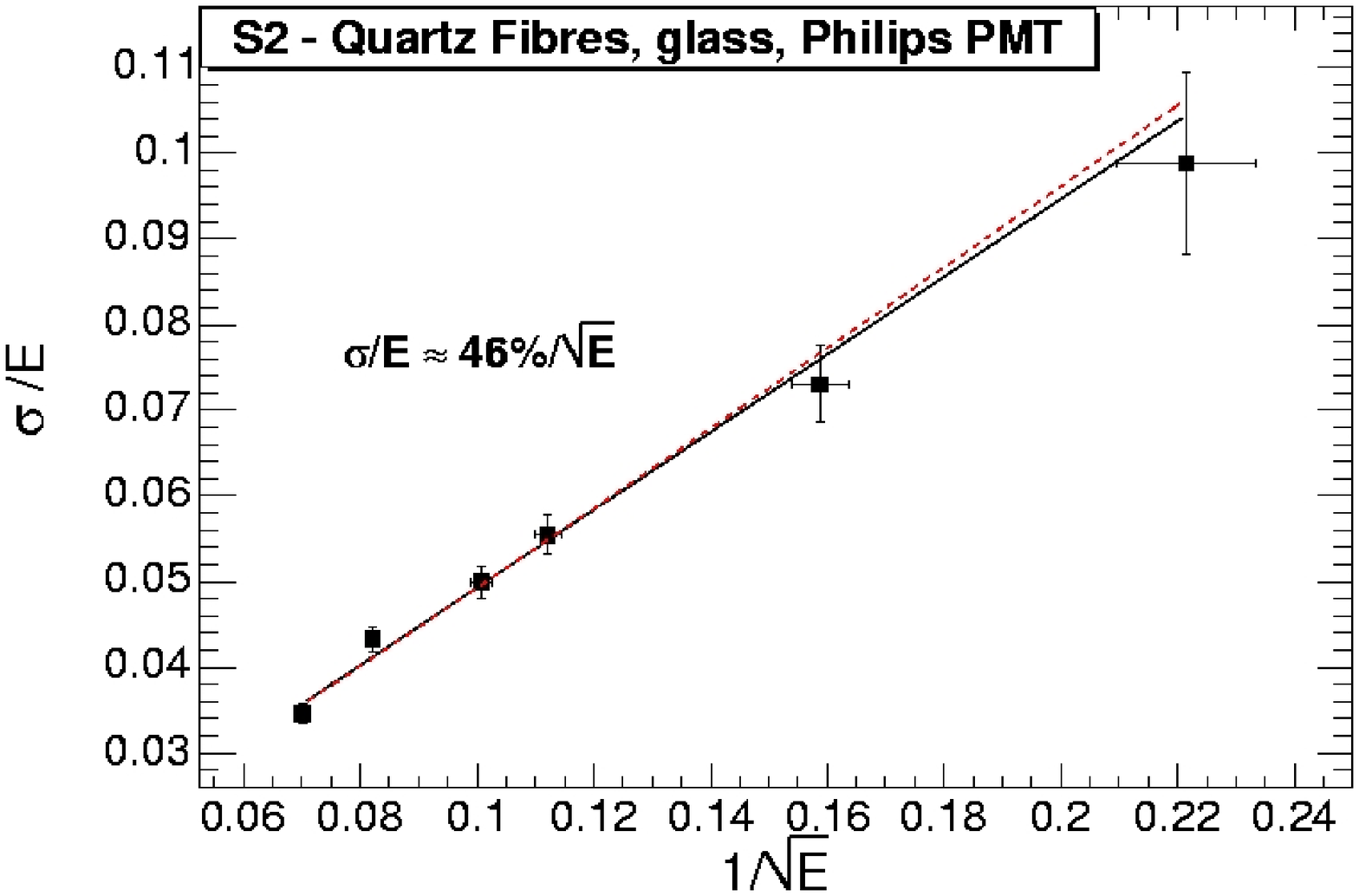} \\
\includegraphics[width=6.8cm]{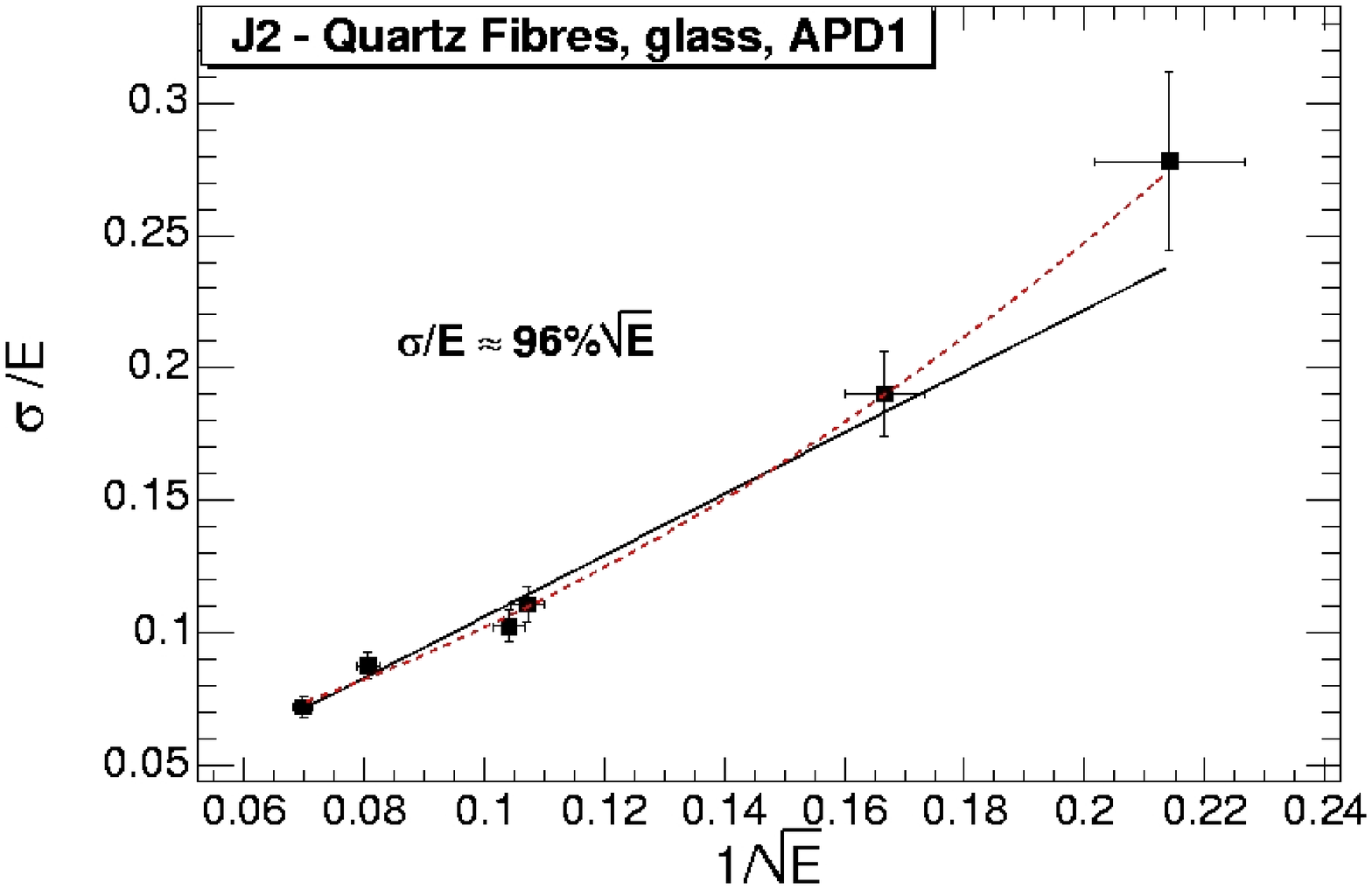}
\includegraphics[width=6.8cm]{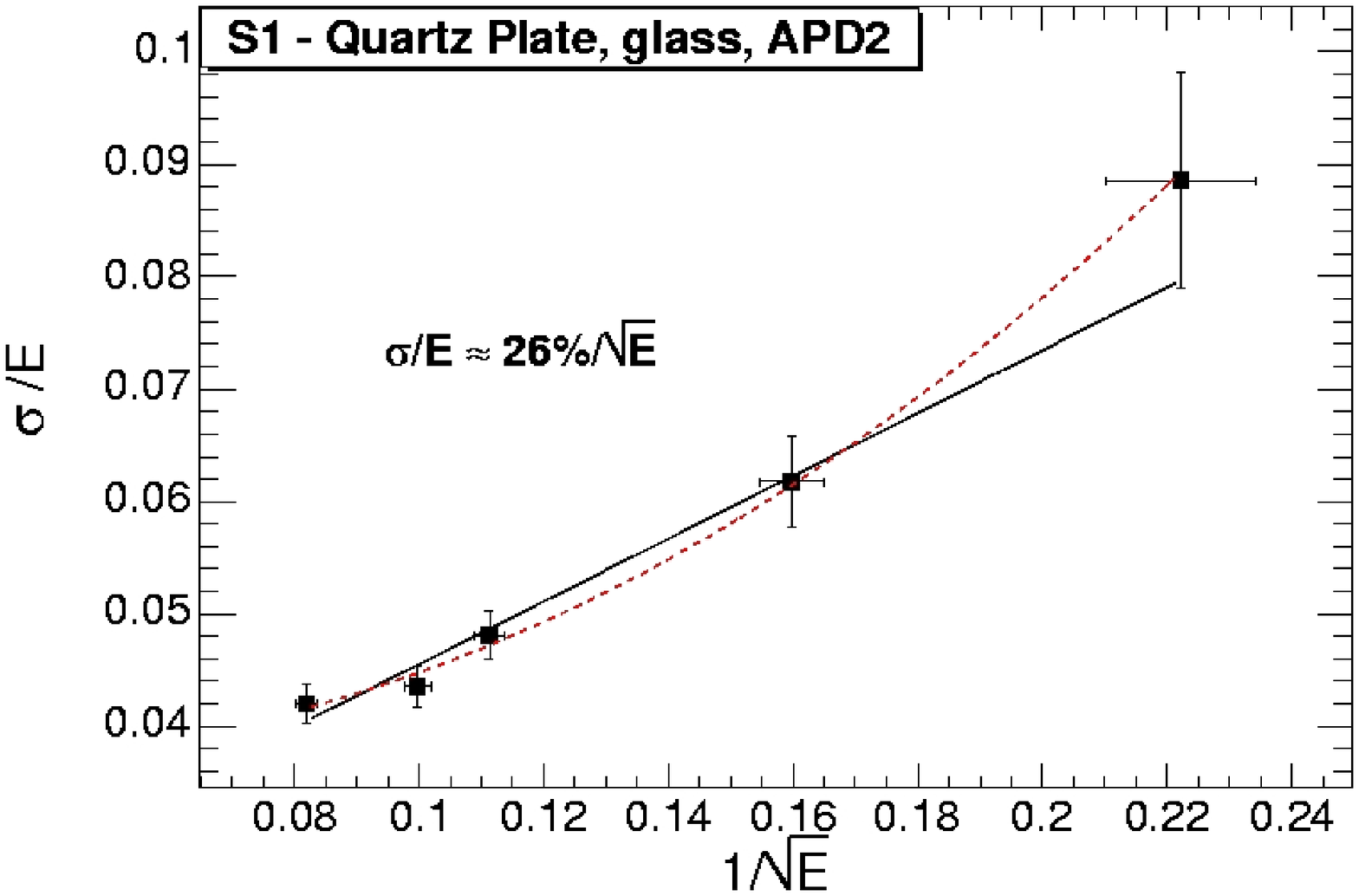} \\
\caption{Energy resolution in sectors: (a) S1 (Philips PMT), 
(b) S2 (Philips PMT), (c) J2 (APD1), (d) S1 (APD2). Two fits are 
shown:  $\sigma/E = p_0 + p_1/\sqrt{E}$ (solid); 
$\sigma/E = p_0 \oplus p_1/\sqrt{E} \oplus p_2/E$ (dashed), with $E$ 
given in GeV. The quoted $\sigma/E$ values are an average between 
both fits.}
\label{fig:energy_resol}
\end{center}
\end{figure} 

Generally, both formulae satisfactorily fit the data (Fig.~\ref{fig:energy_resol}). 
The fit parameters are shown in Table~\ref{tab:linearity_resol}. The first thing to notice 
is that the constant term $p_0$ is close to 0 for all options. The average stochastic term  
$p_1$ is in the range $\sim$ 26\% -- 96\%
and indicates that we can measure the total Pb+Pb electromagnetic energy deposited in CASTOR 
at LHC energies ($\sim$ 40 TeV, according to {\sc hijing}~\cite{hijing}) with a resolution
around 1\%. The readout by avalanche photodiodes leads to the  $p_2$ term, measured 
to be 1.25 GeV and 4.5 GeV for Advanced Photonix APD and Hamamatsu APD, respectively. 
It should be noted that the APDs are very sensitive to both voltage and temperature changes, 
but in this test there was no such stabilization. In Table~\ref{tab:linearity_resol} we summarize 
the fit parameters for both parameterizations and for the four considered configurations.

\begin{table}[H]
\centering
\caption{Energy linearity and resolution of four different configurations of the CASTOR 
calorimeter prototype. For the energy resolution, we quote the parameters for two fits: 
(1) $\sigma/E = p_0 + p_1/\sqrt{E}$, and (2) $\sigma/E = p_0 \oplus p_1/\sqrt{E} \oplus p_2/E$
with $E$ given in GeV.}
\vskip 0.2cm
\label{tab:linearity_resol}
\renewcommand{\arraystretch}{1.5}
\scriptsize{
\begin{tabular}{|l|ccccc|ccc|}\hline\hline
		       & \multicolumn{5}{c}{Resolution} &  \multicolumn{3}{c|}{Linearity}  \\
		       &  fit & $p_0$ & $p_1$ & $p_2$ & $\chi^2/$ndf &  $a$ & $b$ & $\chi^2/$ndf \\
		       &      &  & (GeV$^{1/2}$)  & (GeV) &  &  & (GeV$^{-1}$) &   \\\hline
Quartz Plate (S1, glass) &  &	 & & & & & & \\	
Philips PMT                  & (1) & 0.004 $\pm$ 0.002 & 0.36 $\pm$ 0.02 & & 6.4/4 & 37. $\pm$ 12. & 7.7 $\pm$ 0.2 & 4.2/4 \\
 	                     & (2) & 0.010 $\pm$ 0.004 & 0.38 $\pm$ 0.02 & 0.0 $\pm$ 0.4 &  7.4/3 & & &\\
Adv. Photonix APD       & (1) & 0.017 $\pm$ 0.005  & 0.28 $\pm$ 0.04 & & 2.5/3 & 32.5 $\pm$ 2.4 & 4.4 $\pm$ 0.1 & 2.2/3 \\
                                   & (2) & 0.036 $\pm$ 0.006  & 0.24 $\pm$ 0.04 & 1.2 $\pm$ 0.2 &  6.2/2 & & &\\\hline 
Quartz Fibres (S2, glass) &  &	 & & & & & & \\	
Philips PMT                  & (1) & 0.004 $\pm$ 0.003 & 0.45 $\pm$ 0.04 & & 3.2/4 & 33.6 $\pm$ 9.7 & 4.6 $\pm$ 0.1 & 0.41/4 \\
 	                     & (2) & 0.013 $\pm$ 0.006 & 0.48 $\pm$ 0.02 & 0.0 $\pm$ 0.8 &  3.7/3 & & &\\\hline 
Quartz Fibres (J2, glass) &  &	 & & & & & & \\	
Adv. Photonix APD       & (1) & -0.01 $\pm$ 0.01  & 1.16 $\pm$ 0.13 & & 4.1/4 & 29.8 $\pm$ 3.3 & 1.4 $\pm$ 0.1 & 6.5/4 \\
                                   & (2) & 0.04 $\pm$ 0.02  & 0.82 $\pm$ 0.22 & 4.5 $\pm$ 1.6 &  1.3/3 & & &\\\hline\hline 
\end{tabular}
}
\renewcommand{\arraystretch}{1}
\end{table}


\subsection{Area scanning}
\label{sec:area_scan}

The purpose of the area scanning was to check the uniformity of the calorimeter response, 
affected by electrons hitting points at different places on the sector area, as well as to 
assess the amount of ``edge effects'' and lateral leakage from the calorimeter, leading to 
cross-talk between neighbouring sectors.

For the area scanning of sector S2, connected to the Philips PMT, central points (A-E) as 
well as border points (I-O) have been exposed to electron beam of energy 100 GeV 
(see Fig.~\ref{fig:proto_scheme}). The distributions are symmetric and well described by 
Gaussian fits for the majority of the points. Asymmetric distributions are seen only for 
points closer than $\sim$3 mm to the calorimeter outer edge or sector border.

\begin{figure}[H] 
\begin{center}
\includegraphics[width=13cm]{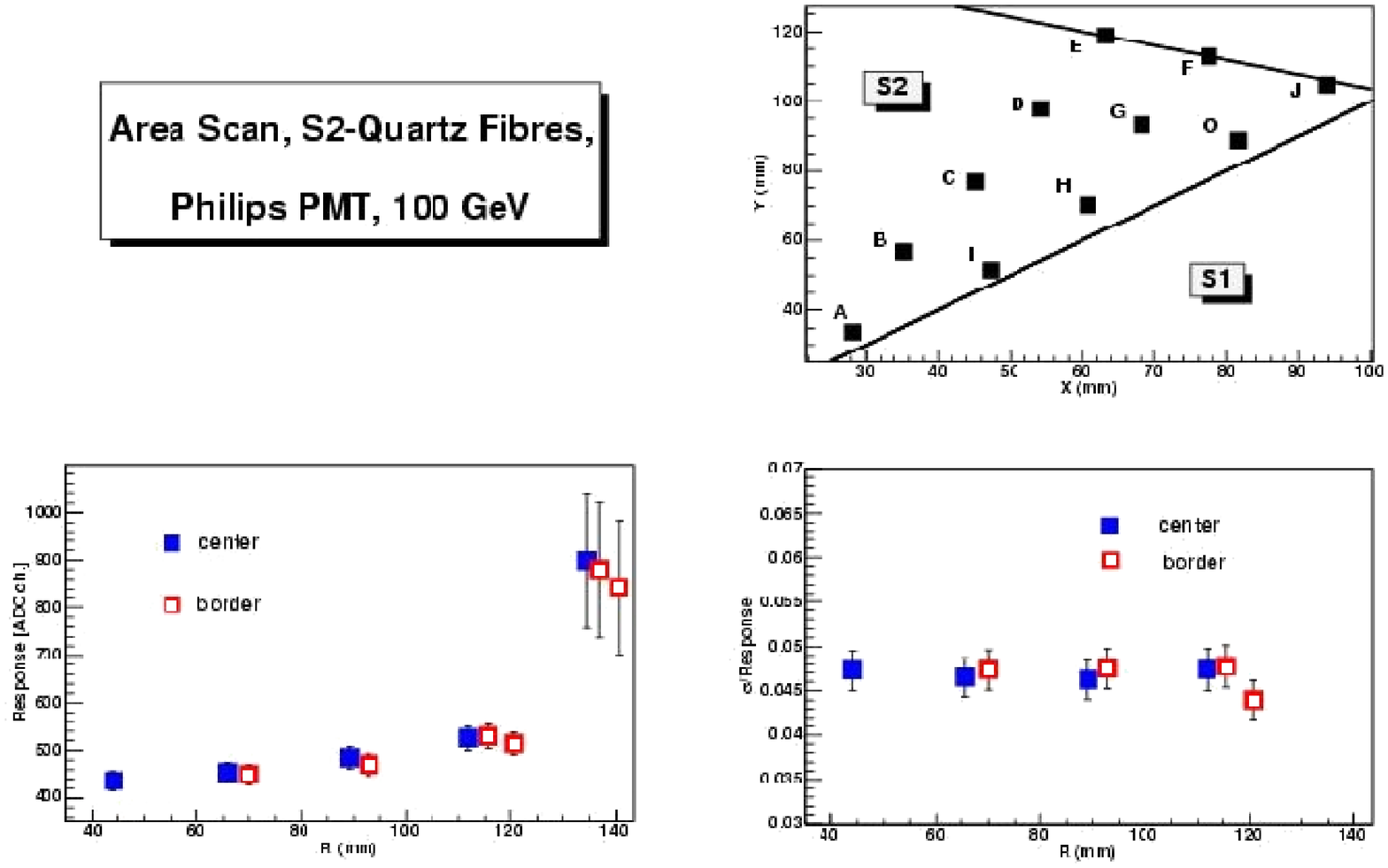}
\caption{Dependence of signal amplitude on the distance $R$ from the 
calorimeter center in sector S2 (Philips PMT). Top: Coordinates of  
the scanned points. Bottom plots: Measured response to 100 GeV 
electrons on central (A-E, filled squares) and border (I-O, hollow squares) points.}
\label{fig:energy_impact}
\end{center}
\end{figure}

Figure~\ref{fig:energy_impact} shows the calorimeter response and relative resolution ($\sigma/E$) 
as a function of the distance $R$ from the calorimeter center, for both central and border points. 
The top plot shows the coordinates of the points, corrected for the beam impact point position. 
It can be seen that points E, F, J practically lie at the upper edge of the calorimeter. The rise of 
the signal amplitudes (bottom left), as well as of the distribution widths with R can be attributed 
to a lateral spread of the beam. For large $R$, a substantial part of the electron beam is outside of 
the calorimeter sector and falls directly onto the light guides. The bottom right plot shows 
that the energy resolution is $\sim$ 4.7\% for 100 GeV electrons and is relatively independent 
of the position of the impact points.

\subsubsection{S1 - S2 cross talk}

\begin{figure}[H] 
\begin{center}
\resizebox{12cm}{!}
{\includegraphics{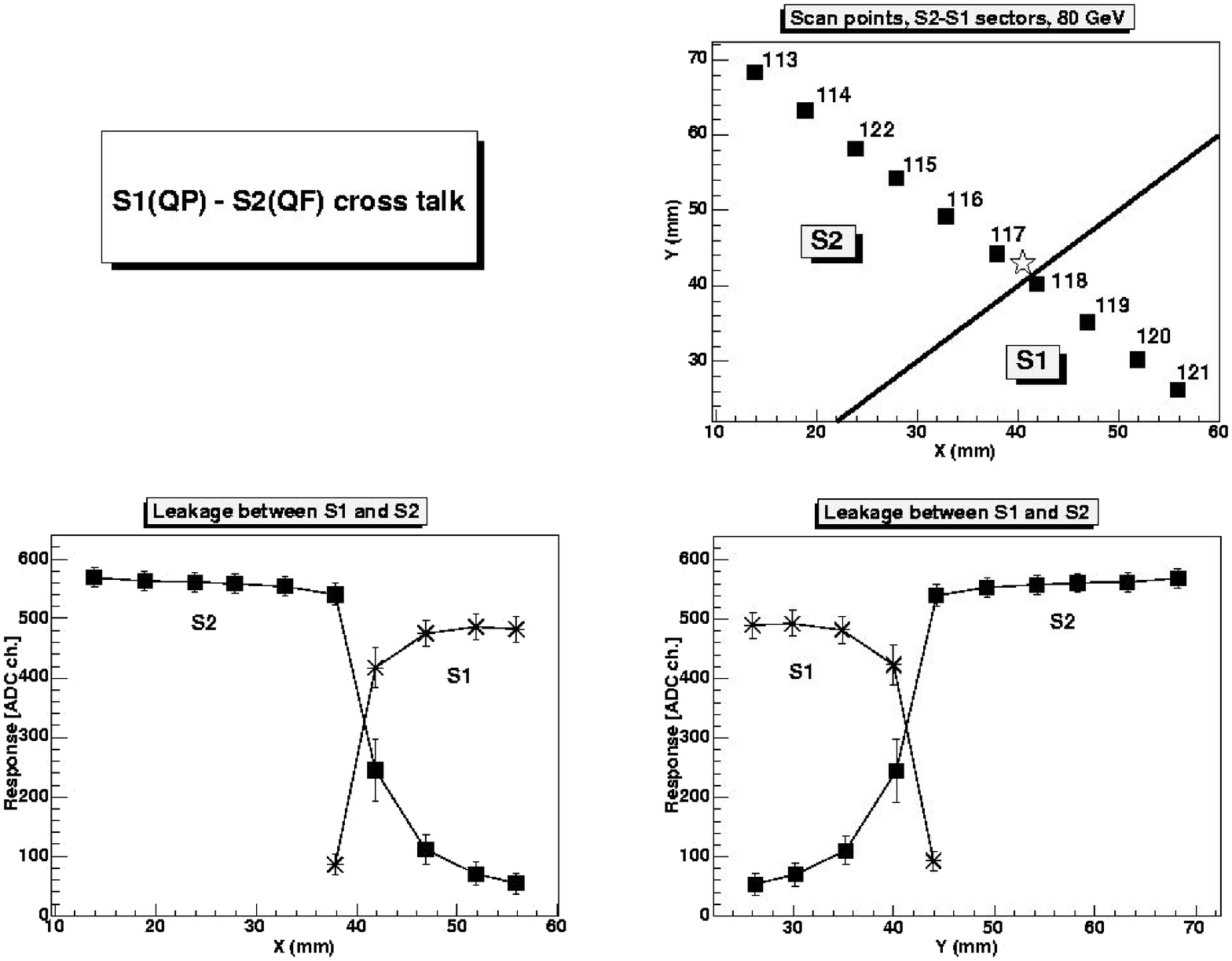}}
\caption{Top: Position of the points in the calorimeter frame, corrected for the beam 
impact points. Bottom: Measured calorimeter response versus coordinates X (left) and Y (right) 
in sectors S1 (APD2) and S2 (Hamamatsu PMT) for several points at distances $\sim$ 2.5-32. mm from 
the sector border.}
\label{fig:impact_point1}
\end{center}
\end{figure}

Ten points, located at distances 2.5-32. mm from the S1/S2 sector border, have been exposed 
to the electron beam of energy 80 GeV. The simultaneous readout of both sectors has been 
done by Advanced Photonix APD and Hamamatsu PMT in S1 and S2, respectively. The upper left 
pad of Figure~\ref{fig:impact_point1} shows the coordinates of the measured points 
in the calorimeter frame, corrected for the beam impact point position. The star symbol 
marks the coordinates of the border point between S1 and S2 sectors, found from the 
dependence of the signal amplitudes on X(Y) coordinates (lower pads). 

The distributions of the signal amplitudes in S2 sector, for points distanced from the sector 
border more than $\sim$ 8 mm, are symmetric (Gaussian) and leakage to S1 sector is negligible. The 
relative energy resolution $\sigma/E$ is of the order $\sim$ 2.9\% for 80 GeV electrons.

\begin{figure}[H] 
\begin{center}
\resizebox{12cm}{!}
{\includegraphics{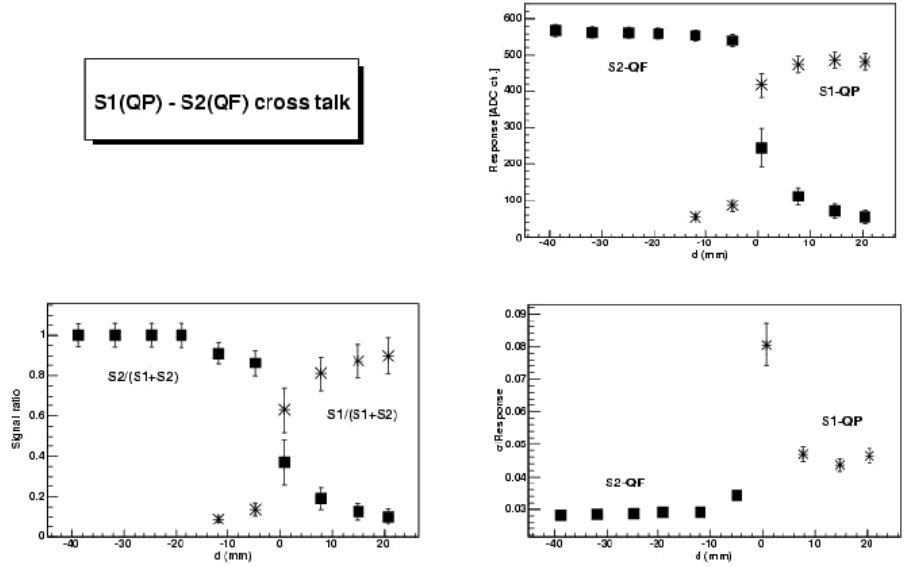}}
\caption{Comparison of the calorimeter response (top right), leakage fraction (bottom left), 
and relative energy resolution, $\sigma$/response, (bottom right) in sectors S1 (APD2) and S2 
(Hamamatsu PMT) for points at different distances $d$ from the sector border.}
\label{fig:impact_point2}
\end{center}
\end{figure}

The dependence of the calorimeter response, leakage fraction and relative energy resolution, 
$\sigma$/response, on the distance $d$ from the sector border, for S1 and S2 sectors are shown in 
Figure~\ref{fig:impact_point2}. 

Both the light output and energy resolution are a little better 
for S2 sector, connected to Hamamatsu PMT ($\sigma/E$ $\sim$ 2.9\%), than for S1 sector, connected to 
Advanced Photonix APD ($\sigma/E$ $\sim$ 4.5\%). This is expected since there is more light collected 
by the PMT as compared to the APD: area(PMT)/area(APD) = 1.55.


\subsubsection{Comparison of J1, J2 and S1 sectors}

For comparison of the uniformity of calorimeter response, several points located at different 
places on the sectors have been exposed to the electron beam of 80 GeV energy. The points 
(A-E) at the middle of J1, J2 and S1 sectors and points (4-8) at the border of S1 sector have been 
studied (see Figure~\ref{fig:proto_scheme}). All sectors have been connected to Hamamatsu PMT. 
Gaussian distributions of signal amplitudes in the middle of the sectors and 
asymmetric distributions close to the sector border (points 4-8) and sometimes also close 
to the inner (point A) and outer (point E) calorimeter edge in J1 sector are observed. 
The beam profile correction (aiming at selecting the central core of the impinging beam) 
reduces the asymmetry. 

\begin{figure}[H] 
\begin{center}
\includegraphics[width=6.2cm]{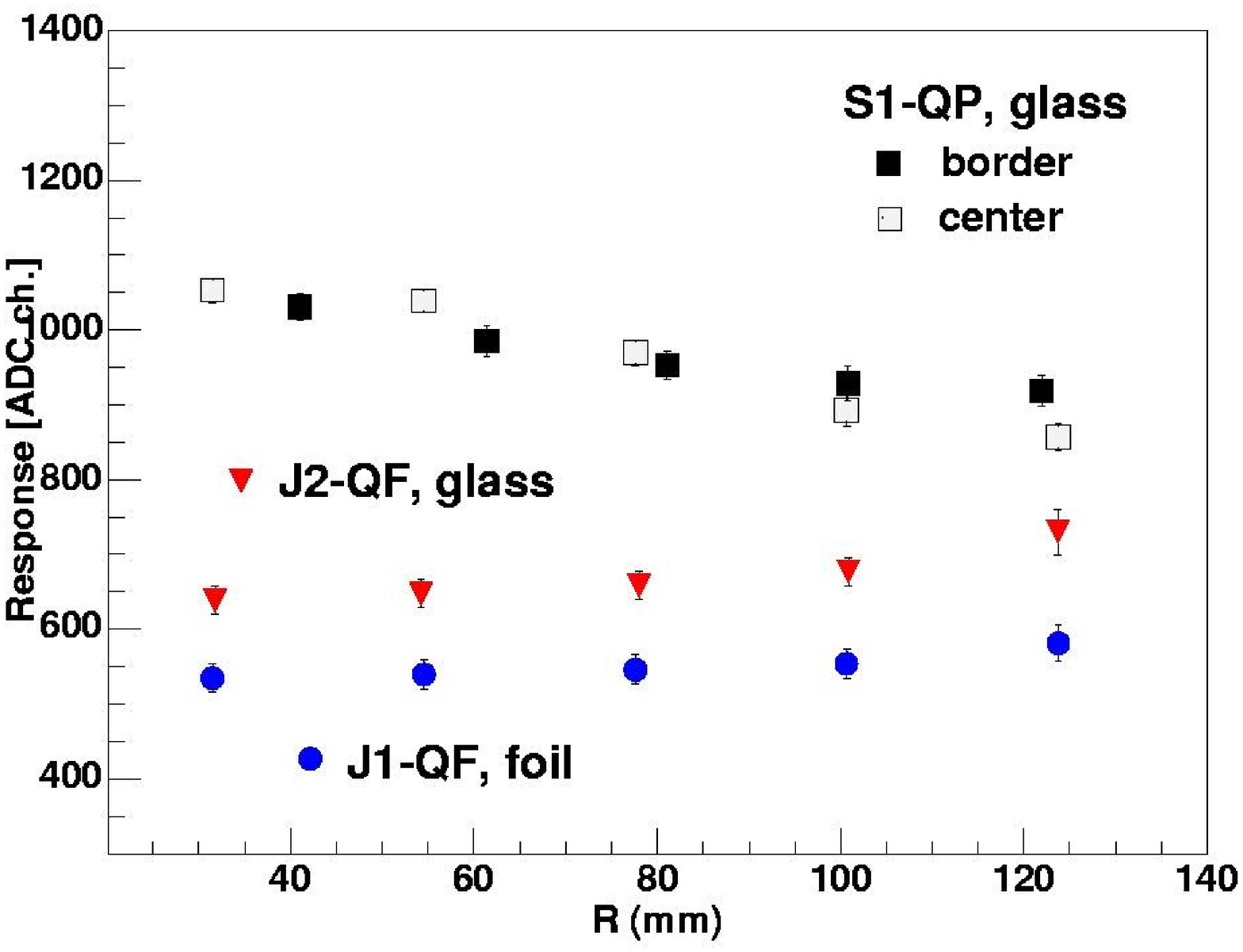}
\includegraphics[width=6.5cm,height=4.75cm]{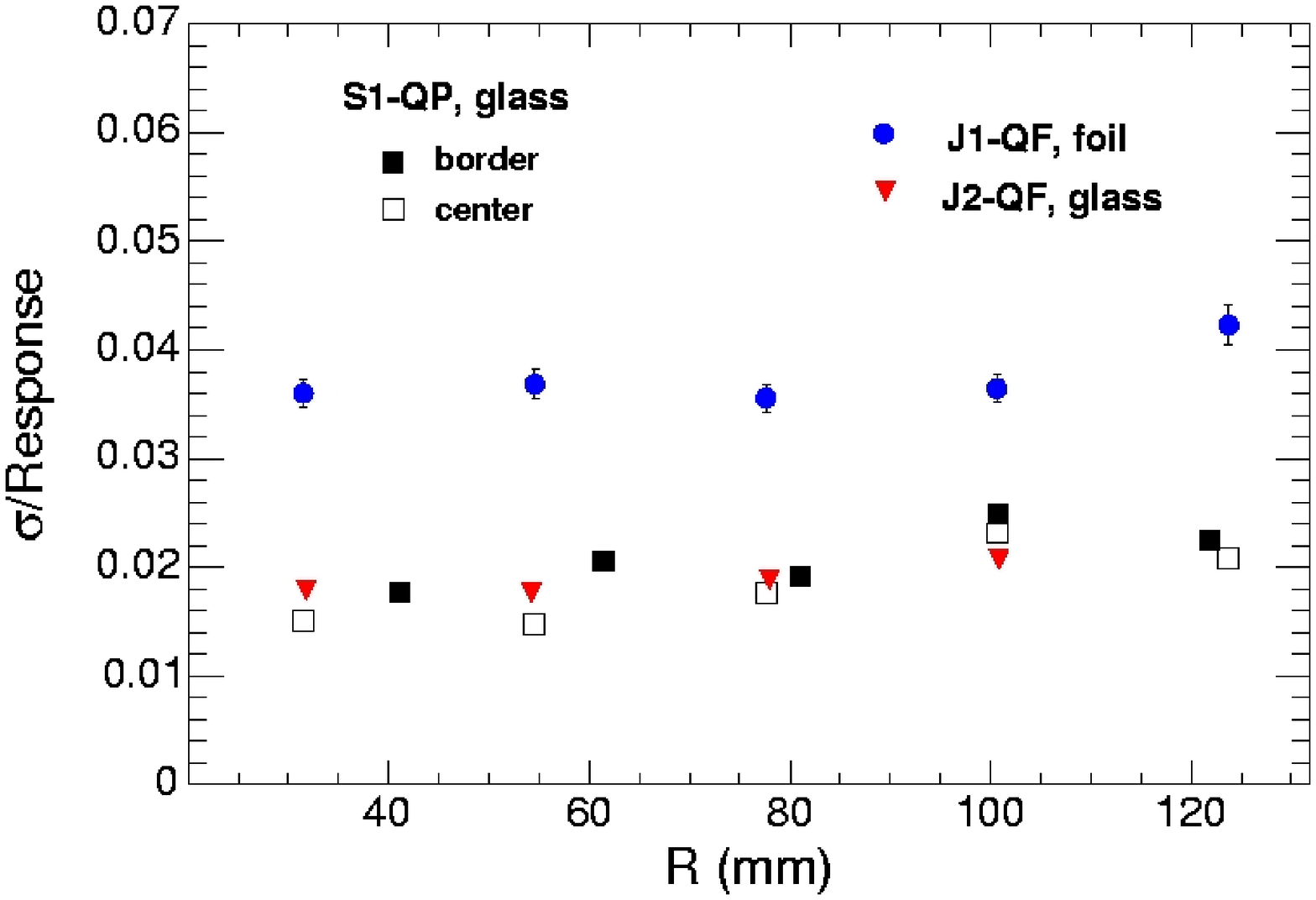}
\caption{Comparison of calorimeter response (left) and resolution (right) to 80 GeV 
electrons for several impact points (A-E) of J2, J1 and S1 sectors, readout with Hamamatsu PMTs.}
\label{fig:j1j2s1_comparison}
\end{center}
\end{figure}

Comparison of light output and relative energy resolution for all options studied is shown 
in Figure~\ref{fig:j1j2s1_comparison}. Light output is highest in the S1 (QP-glass) sector and it 
is practically the same for the central and border points. It depends weakly on the distance $R$ of
the impact point. For S1, a weak decrease and for J1 and J2 sectors a weak increase of the calorimeter 
response with distance R from the calorimeter center are observed. 
The relative energy resolution is almost independent of the position of the impact point 
and it is $\sim$ 1.5-2.5 \% for S1 (QP-glass) and J2 (QF-glass) sectors and $\sim$ 3.5-4.0 \% for J2 
(QF-foil) for 80 GeV electrons. 


\section{Summary}

We have presented a comparative study of the performances of the first prototype of the 
CASTOR quartz-tungsten calorimeter of the CMS experiment using different detector 
configurations. {\sc geant}-based MC simulations have been employed to determine the \v{C}erenkov 
light efficiency of different types of air-core light guides and reflectors. Different sectors of the 
calorimeter have been setup with various quartz active materials and with different photodetector
devices (PMTs, APDs). Electron beam tests, carried out at CERN SPS in 2003, have been used to 
analyze the calorimeter linearity and resolution as a function of energy and impact point. 
The main results obtained can be summarized as follows:

\begin{description}
\item 1. Comparison between the calorimeter response using a single quartz plate or using a 
quartz-fibre bundle indicates that:
\item (a) Good energy linearity is observed for both active medium options (Fig.~\ref{fig:linearity}).
\item (b) The Q-plate gives more light output than equal thickness Q-fibres (Fig.~\ref{fig:j1j2s1_comparison}).
\item (c) The relative energy resolution is similar for quartz plates and quartz fibres (Fig.~\ref{fig:energy_resol}). 
When readout with the same Hamamatsu PMT (S1, S2 sectors), we found $\sim$2\% energy resolution 
for 80 GeV electrons (Fig.~\ref{fig:j1j2s1_comparison}). 
\item (d) The constant term  $p_0$ of the energy resolution, that limits performance at high energies, 
is less than 1\% in both options for the same Philips PMT and glass reflector (Fig.~\ref{fig:energy_resol}).
The stochastic term  $p_1$ is $\sim$36 \% and $\sim$46\% for quartz plates and quartz fibres, 
respectively (Table~\ref{tab:linearity_resol}).

\item 2. Avalanche-photodiodes (APDs) appear to be a working option for the photodetectors,
although they still need more investigation (radiation-hardness, cooling and voltage stabilization tests). 

\item 3. The relative energy resolution is weakly dependent on the position of the impact point 
(Fig.~\ref{fig:j1j2s1_comparison}). Leakage (cross-talk) between sectors is negligible for impact 
points separated more than 8 mm from the sector border. Only, electrons impinging less than 
3 mm from the detector edge show a degraded energy response and worse resolution.
\vspace{1mm}

\item 4. The shape of the light guide is determined by tree parameters: (i) the type of quartz fiber 
(NA number), (ii) the maximum efficiency and uniformity of response, and (iii) the available space for 
the size of a calorimeter. The aim is to simultaneously achieve optimum efficiency and uniformity 
of light transmission within the realistically available space. From the analysis of the MC simulations 
we come to the conclusion that the above requirements are best satisfied with $lm$ = 0 and $lg$ = 2 
for NA = 0.37 and 0.48.

\item 5. The light output is a little higher for the light-guides with glass reflector compared 
to those that use HF-foil, for the same photodetector (Hamamatsu PMT,  Fig.~\ref{fig:j1j2s1_comparison}). 
This is understood, since the HF reflecting foil is designed to cut \v{C}erenkov light 
with $\lambda$ $<$ 400 nm, where the light output is greater. However, the HF-reflector foil has 
higher efficiency in the region $\lambda>$ 400 nm than the glass mirror 
(Table~\ref{tab:transmittance}). 
\end{description}

In summary, this study suggests that equipping the CASTOR calorimeter with quartz-plates 
as active material, APDs as photodetector devices (with temperature and voltage stabilization), 
and light-guides with foil reflector is a promising option, although the final configuration 
would benefit from further (detailed) investigation to take into account the experimental 
conditions that will be encountered in the forward rapidity region of CMS. A beam test of the 
second prototype was carried out in 2004 and the results are reported elsewhere~\cite{bt-2004}.


\section{Acknowledgments}

We wish to thank R.~Wigmans and N.~Akchurin for assistance in the early stage of the beam
test. This work is supported in part by the Secretariat for Research of the University of
Athens and the Polish State Committee for Scientific Research (KBN) 
SPUB-M nr. 620/E-77/SPB/CERN/P-03/DWM 51/2004-2006. D.d'E. acknowledges support
from the 6th EU Framework Programme (contract MEIF-CT-2005-025073).


\end{document}